\documentclass[11pt]{article}
\usepackage{subcaption, booktabs, siunitx, adjustbox, bm}
\usepackage{mathrsfs}
\usepackage{algorithm}
\usepackage{algpseudocode}
\usepackage{amsmath, amssymb, mathtools, nicefrac}
\usepackage{cite}

\usepackage{lipsum}
\usepackage{amsfonts}
\usepackage{graphicx}
\usepackage{epstopdf}
\ifpdf
  \DeclareGraphicsExtensions{.eps,.pdf,.png,.jpg}
\else
  \DeclareGraphicsExtensions{.eps}
\fi
\newcolumntype{C}{>{\centering\arraybackslash}m{2cm}}
\newcolumntype{N}{>{\centering\arraybackslash}m{2.5cm}}

\newcommand{\x}{\textbf{x}}

\newtheorem{example}{Example}
\newtheorem{theorem}{Theorem}
\newtheorem{definition}{Definition}
\newtheorem{lemma}{Lemma}
\newtheorem{remark}{Remark}

\newtheorem{corollary}{Corollary}
\newcommand{\qed}{\hfill $\blacksquare$}

\title{\bf\Large Bounds on the Excess Minimum Risk via Generalized Information Divergence Measures\thanks{This work was supported in part by the Natural Sciences and Engineering Research Council (NSERC) of Canada. This work was presented in part at the {\em International Symposium on Information Theory and Its Applications}, Taipei, Taiwan, November 2024~\cite{omanwar2024bounding}.

The authors are with the Department of Mathematics and Statistics, Queen's University, Kingston, Ontario, Canada
  (\{a.omanwar,fa,tamas.linder\}@queensu.ca).}}

\author{Ananya Omanwar
\and Fady Alajaji
\and Tam\'as Linder
}

\usepackage{amsopn}

\allowdisplaybreaks

\begin{document}
\date{}
\maketitle

\abstract{Given finite-dimensional random vectors $Y$, $X$, and $Z$ that form a Markov chain in that order ($Y \to X \to Z$), we derive upper bounds on the excess minimum risk using generalized information divergence measures. Here, $Y$ is a target vector to be estimated from an observed feature vector $X$ or its stochastically degraded version $Z$. The excess minimum risk is defined as the difference between the minimum expected loss in estimating $Y$ from $X$ and from $Z$. We present a family of bounds that generalize the mutual information based bound of Gy\"orfi~{\em et al.} (2023), using the R\'enyi and $\alpha$-Jensen-Shannon divergences, as well as Sibson's mutual information. Our bounds are similar to those developed by Modak~{\em et al.} (2021) and Aminian~{\em et al.} (2024) for the generalization error of learning algorithms. However, unlike these works, our bounds do not require the sub-Gaussian parameter to be constant and therefore apply to a broader class of joint distributions over $Y$, $X$, and~$Z$. We also provide numerical examples under both constant and non-constant sub-Gaussianity assumptions, illustrating that our generalized divergence based bounds can be tighter than the one based on mutual information for certain regimes of the parameter $\alpha$.
 }
 

\section{Introduction}

\smallskip

The excess minimum risk in statistical inference quantifies the difference between the minimum expected loss attained by estimating a (target) hidden random vector from a feature (observed) random vector and the minimum expected loss incurred by estimating the hidden vector from a stochastically degraded version of the feature vector. The aim of this work is to derive upper bounds on the excess minimum risk in terms of generalized information divergence measures such as the R\'enyi divergence~\cite{renyi}, the $\alpha$-Jensen-Shannon divergence~\cite{AJS1,AJS2} and the Sibson mutual information~\cite{ASM1,Csiszar}. 

\medskip

Recently, several bounds of this nature, expressed in terms of information-theoretic measures, have appeared in the literature, including~\cite{Xu2,bu20,Esposito,Esposito1,esposito21,Modak,ji21,Xu1,Gyorfi,hafez-kolahi23,aminian24,Aminian} among others. Most of these works have focused on the (expected) generalization error of learning algorithms. In~\cite{Xu2}, Xu and Raginsky established bounds on the generalization error in terms of Shannon's mutual information between the (input) training dataset and the (output) hypothesis; these bounds are tightened in~\cite{bu20} by using the mutual information between individual data samples and the hypothesis. In~\cite{Modak}, Modak~{\em et al.} extend these works by obtaining upper bounds on the generalization error in terms of the R\'enyi divergence, employing the variational characterization of the R\'enyi divergence~\cite{DVRformula,DVRformula2,DVRformula3}. The authors also derive bounds on the probability of generalization error via R\'enyi's divergence, which recover the bounds of Esposito~{\em et al.}~\cite{Esposito1} (see also~\cite{Esposito,esposito21} for bounds expressed in terms of the $f$-divergence~\cite{csiszar-fdiv1}). More recently, Aminian~{\em et al.}~\cite{Aminian} obtained a family of bounds on the generalization error and excess risk applicable to supervised learning settings using a so-called ``auxiliary distribution method.'' In particular, they derive new bounds based on the $\alpha$-Jensen-Shannon and $\alpha$-R\'enyi mutual information measures. Here, both measures are defined via divergences between a joint distribution and a product of its marginals: the former using the Jensen-Shannon divergence of weight~$\alpha$~\cite[Eq.~(4.1)]{lin91} (which is always finite), and the latter using the R\'enyi divergence of order~$\alpha$. In addition to information-theoretic approaches, generalization bounds based on PAC-Bayesian theory~\cite{mcallester99,alquier16}, particularly those involving $f$-divergences and R\'enyi-type divergences, have been actively studied. Separately, generalization bounds based on the Wasserstein distance~\cite{lopez20} have also been established as an alternative approach based on optimal transport techniques. Other works on the analysis of generalization error include~\cite{ji21,welfert23} for deep learning generative adversarial networks~\cite{gan} and~\cite{aminian24} for the Gibbs algorithm (see also the extensive lists of references therein).

\medskip

In this work, we focus on the excess minimum risk in statistical inference. Our motivation is to generalize the results of Gy\"orfi~{\em et al.}~\cite{Gyorfi}, who derived a  mutual information based upper bound that applies to a broad class of loss functions under standard sub-Gaussianity assumptions. Related but distinct work includes~\cite{Xu1,hafez-kolahi23}, where information-theoretic bounds on excess risk are developed in a Bayesian learning framework involving training data. We extend the bound in~\cite{Gyorfi} by introducing a family of bounds parameterized by an order $\alpha \in (0,1)$, based on generalized information divergence measures—namely, the R\'enyi divergence, the $\alpha$-Jensen-Shannon divergence and the Sibson mutual information. For the R\'enyi divergence based bounds, we adopt an approach similar to~\cite{Modak}, leveraging its variational representation. For the bounds involving the $\alpha$-Jensen-Shannon divergence and the Sibson mutual information, we follow the methodology of~\cite{Aminian}, employing the auxiliary distribution method together with the variational representation of the Kullback--Leibler (KL) divergence~\cite{DV}. Unlike~\cite{Modak} and~\cite{Aminian}, where the sub-Gaussian parameter is assumed to be constant, our setup allows this parameter to depend on the (target) random vector being estimated. This makes our bounds applicable to a broader class of joint distributions over the random vectors involved.

\medskip

Our problem of bounding the excess minimum risk is closely related to recent work on generalization error in learning theory. In both settings, the goal is to understand how much performance is lost when a target variable is estimated from a less informative or transformed version of the input. In learning theory, this is often studied through generalization bounds, which compare the performance of a learned predictor on training and test data. As already stated, several recent works have used information-theoretic tools—such as mutual information and its generalizations—to bound the generalization error (e.g.,~\cite{Xu2,Modak,Aminian}). Although these works focus on algorithm-dependent error, the structure of the bounds is similar to ours. Our bounds, instead, are on the excess minimum risk, which compares the best possible performance using full observations versus using degraded ones. Still, both approaches rely on similar tools, including variational characterizations and divergence measures. In this sense, our work takes a different but related approach by studying the basic limits of inference, rather than how well a particular algorithm performs.

\medskip

This paper is organized as follows. In Section~\ref{sec:Preliminaries}, we provide preliminary definitions and introduce the statistical inference problem. In Section~\ref{sec:Main}, we establish a family of upper bounds on the excess minimum risk, expressed in terms of the R\'enyi divergence, the $\alpha$-Jensen-Shannon divergence and the Sibson mutual information, all parameterized by the order $\alpha \in (0,1)$.We also present several numerical examples, including cases with both constant and non-constant sub-Gaussian parameters, all of which demonstrate that the proposed bounds are tighter than the mutual information bound for a range of values of $\alpha$. In Section~\ref{sec:Conclusion}, we provide concluding remarks and suggest directions for future work.

\section{Preliminaries}\label{sec:Preliminaries}
\smallskip
\subsection{Problem Setup}\label{subsec:problem_setup}
Consider a random vector $Y \in \mathbb{R}^p$, $p\geq 1$, that is to be estimated (predicted) from a random observation vector $X$ taking values in $\mathbb{R}^q$, $q\geq 1$. Given a measurable estimator (predictor) $f:\mathbb{R}^q\rightarrow\mathbb{R}^p$ and a loss function $l:\mathbb{R}^p\times\mathbb{R}^p\rightarrow\mathbb{R}^{+}$, the loss (risk) realized in estimating $Y$ by $f(X)$ is given by $l(Y,f(X))$. The minimum expected risk in predicting $Y$ from $X$ is defined by
\begin{equation}\label{eq:MER}
    L_{l}^{*}(Y|X) = \inf_{f:\mathbb{R}^q\rightarrow\mathbb{R}^p}\mathbb{E}[l(Y,f(X))] 
\end{equation}
where the infimum is over all measurable $f$.

\medskip

We also consider another random observation vector $Z$ that is a random transformation or stochastically degraded version of $X$, obtained for example by observing $X$ through a noisy channel. Here $Z$ takes values in $\mathbb{R}^{r}$, $r\geq1$, and $Y$, $X$ and $Z$ form a Markov chain in this order, which we denote as $Y\rightarrow X\rightarrow Z$. We similarly define the minimum expected risk in predicting $Y$ from $Z$ as
\begin{equation}\label{eq: MER with Z}
    L_{l}^{*}(Y|Z) = \inf_{g:\mathbb{R}^{r}\rightarrow\mathbb{R}^p}\mathbb{E}[l(Y,g(Z))], 
\end{equation}
where the infimum is over all measurable predictors $g$. With the notation introduced above, we define the excess minimum risk as the difference $L_{l}^{*}(Y|Z) - L_{l}^{*}(Y|X)$, which is always non-negative due to the Markov chain condition $Y\rightarrow X\rightarrow Z$ (e.g., see the data processing inequality for expected risk in~\cite[Lemma~1]{Xu1}).  
Our objective is to establish upper bounds to this difference using generalized information divergence measures. 

\smallskip
In~\cite{Gyorfi}, the random vector $Z$ is taken as $T(X)$, a transformation of random vector $X$, where $T:\mathbb{R}^{p}\rightarrow\mathbb{R}^{r}$ is measurable. The authors derive bounds on the excess minimum risk using Shannon's mutual information. Here, we generalize these bounds by employing a family of information‐divergence measures of order $\alpha\in(0,1)$, which recover Shannon's mutual information in the limits $\alpha\to0$ or $\alpha\to1$. Furthermore, we use an arbitrary random vector $Z$, as the degraded version of the observation $X$ instead of $T(X)$. We also provide examples where the various generalized information divergence based bounds perform better than the mutual information based bounds of~\cite{Gyorfi} 
for a certain range $\alpha$.

\medskip

We next state some definitions that we will invoke in deriving our results.

\subsection{Definitions}\label{subsec:Definitions}

Consider two arbitrary jointly distributed random variables $U$ and $V$ defined on the same probability space $(\Omega,\mathcal{M})$ and taking values in $\mathcal{U}$ and $\mathcal{V}$, respectively. Let $P_U$ and $P_V$ be the marginal distributions of $U$ and $V$, respectively, and $P_{U,V}$ be a joint distribution over $\mathcal{U}\times\mathcal{V}$. We first provide definitions for the R\'enyi divergence based measures.

\begin{definition}\cite{renyi,Erven}
    The R\'enyi divergence of order $\alpha\in(0,\infty)$,  $\alpha\neq 1$, between two probability measures $P_U$ and $P_V$ is denoted by $D_{\alpha}(P_U\| P_V)$ and defined as follows. Let $\nu$ be a sigma-finite positive measure such that  $P_U$ and $P_V$ are absolutely continuous with respect to $\nu$, written as $P_U,P_V \ll \nu$, with Radon–Nikodym derivatives $\frac{dP_U}{d\nu} = p_U$ and $\frac{dP_V}{d\nu} = p_V$, respectively. Then
    \begin{align*}\label{RD}
    &D_{\alpha}(P_U\| P_V)  =  \left\{
    \begin{array}{ll}
      \frac{1}{\alpha-1}\log\left[\int (p_{U})^{\alpha}(p_{V})^{1-\alpha}d\nu\right] & \mbox{ if } 0 < \alpha <1 \mbox{ or } \alpha >1  \mbox{ and }  P_U \ll P_V \\
      +\infty & \mbox{ if } \alpha>1 \mbox{ and } P_U\ll P_V. \\
    \end{array} \right  .        
    \end{align*}    
\end{definition}

\smallskip

\begin{definition}
    The conditional R\'enyi divergence of order $\alpha$ between the conditional distributions $P_{V|U}$ and $Q_{V|U}$ given $P_U$ is denoted by $D_{\alpha}(P_{V|U}\| Q_{V|U} |P_U)$ and given by
    \begin{equation}\label{CondRD}
       D_{\alpha}(P_{V|U}\| Q_{V|U} |P_U) = \mathbb{E}_{P_U}\big[D_{\alpha}(P_{V|U}(\,\cdot\,|U)\| Q_{V|U}(\,\cdot\,|U)))\big],
    \end{equation}
     where $\mathbb{E}_{P_U}[\,\cdot\,]$ denotes expectation with respect to distribution $P_U$. 
\end{definition}

Note that the above definition of conditional R\'enyi divergence {\em differs} from the standard one, which is given as
$D_\alpha(P_{V|U} P_U \|Q_{V|U} P_U)$, e.g., see~\cite[Definition~3]{verdu15}. However as $\alpha \to 1$, both notions of the conditional R\'enyi divergence recover the conditional KL divergence, which is 
$$D_\text{KL}(P_{V|U}\| Q_{V|U} |P_U)= D_\text{KL}(P_{V|U} P_U \|Q_{V|U} P_U) = \mathbb{E}_{P_U}\left[\int p_{V|U}\log\left[\frac{p_{V|U}}{q_{V|U}}\right]dv\right].$$
\smallskip 

 We next provide the definitions for the $\alpha$-Jensen-Shannon divergence based measures.

\begin{definition}~\cite{AJS1,AJS2}\label{AJSDiv}
    The $\alpha$-Jensen-Shannon divergence for $\alpha\in(0,1)$ between two probability measures $P_U$ and $P_V$ on a measurable space $(\Omega,\mathcal{M})$, is denoted by $JS_{\alpha}(P_U\| P_V)$ and given by
    \begin{equation}
        JS_{\alpha}(P_U \|P_V) = \alpha D_{\text{KL}}(P_U\| \alpha P_U + (1-\alpha)P_V) + (1-\alpha)D_{\text{KL}}(P_V\| \alpha P_U + (1-\alpha)P_V),
    \end{equation}
    where $D_{\text{KL}}(\cdot\|\cdot)$ is the KL divergence.
\end{definition}

\begin{definition}
    The conditional $\alpha$-Jensen-Shannon divergence between the conditional distributions $P_{V|U}$ and $Q_{V|U}$ given $P_U$ is denoted by $JS_{\alpha}(P_{V|U}\| Q_{V|U} |P_U)$ and given by
    \begin{equation}\label{AJSCond}
         JS_{\alpha}(P_{V|U}\| Q_{V|U} |P_U) =  \mathbb{E}_{P_U}\big[JS_{\alpha}(P_{V|U}(\,\cdot\,|U)\| Q_{V|U}(\,\cdot\,|U))\big],
    \end{equation}
     where $\mathbb{E}_{P_U}[\,\cdot\,]$ denotes expectation with respect to distribution $P_U$.
\end{definition}

\smallskip

\noindent We herein define the Sibson mutual information (of order $\alpha$).

\begin{definition}\cite{ASM1,Csiszar}\label{ASibM1}
Let $\alpha \in (0,1) \cup (1, \infty)$. The Sibson mutual information of order $\alpha$ between $U$ and $V$ is denoted by $I_{\alpha}^{S}(U;V)$ and given by
\begin{equation}
    I_{\alpha}^{S}(U;V) = \min_{Q_U \in \mathcal{P}(\mathcal{U})} D_{\alpha}(P_{U,V} \| Q_U P_V),
\end{equation}
where $\mathcal{P}(\mathcal{U})$ denotes the set of probability distributions on $\mathcal{U}$.
\end{definition}

\smallskip

It is known that $D_{\alpha}(P_{U,V} \| Q_U P_V)$ is convex in $Q_U$~\cite{Erven}, which allows for a closed-form expression for the minimizer and, consequently, for the Sibson mutual information $I_{\alpha}^{S}(U;V)$~\cite{Esposito1,Esposito2}. Let $U^*$ denote a random variable whose distribution achieves the minimum, with corresponding distribution $P_{U^*}$. Then, the Sibson mutual information of order $\alpha$ can equivalently be written as:

\begin{definition}\cite{Esposito1}\label{ASibM2}
Let $\nu$ be a sigma-finite positive measure such that $P_{U,V}$ and $P_U P_V$ are absolutely continuous with respect to $\nu \times \nu$, written as $P_{U,V}, P_U P_V \ll \nu \times \nu$, with Radon--Nikodym derivatives(densities) $\frac{dP_{U,V}}{d(\nu \times \nu)} = p_{U,V}$ and $\frac{d(P_U P_V)}{d(\nu \times \nu)} = p_U p_V$, respectively. For $\alpha \in (0,1) \cup (1,\infty)$, the Sibson mutual information of order $\alpha$ between $U$ and $V$ can be written as:
\begin{equation}\label{ASibMd2}
    I_{\alpha}^{S}(U;V) = D_{\alpha}(P_{U,V} \| P_{U^{*}} P_V),
\end{equation}
where the distribution $P_{U^{*}}$ has density
\begin{equation}\label{ASibMd2prime}
    p_{U^{*}}(u) = \frac{dP_{U^*}}{d\nu}(u) = \frac{\left(\displaystyle\int \left(\frac{p_{U,V}(u,v)}{p_U(u) p_V(v)}\right)^\alpha p_V(v)\, dv\right)^{\frac{1}{\alpha}}}{\displaystyle\int \left(\int \left(\frac{p_{U,V}(u',v')}{p_U(u') p_V(v')}\right)^\alpha p_V(v')\, dv'\right)^{\frac{1}{\alpha}} p_U(u')\, du'} \, p_U(u).
\end{equation}
\end{definition}

\smallskip

\begin{remark}\label{AsibMRem}
From Definition~\ref{ASibM2}, we note that the Sibson mutual information of order $\alpha$ is a functional of the distributions $P_{U,V}$ and $P_{U^*}$. Hence, from this point onward, we denote with a slight abuse of notation the Sibson mutual information of order $\alpha$ between $U$ and $V$ by $I_{\alpha}^{S}(P_{U,V}, P_{U^*})$.
\end{remark}

\smallskip

We end this section with the definitions of the sub-Gaussian and conditional sub-Gaussian properties.

\begin{definition}
    A real random variable $U$ with finite expectation is said to be $\sigma^2$-sub-Gaussian for some $\sigma^2>0$ if 
    \begin{equation}\label{SGC}
        \log\mathbb{E}[e^{\lambda(U-\mathbb{E}[U])}]\leq \frac{\sigma^2\lambda^2}{2}
    \end{equation}
    for all $\lambda\in\mathbb{R}$.
\end{definition}

\begin{definition}
    A real random variable $U$ is said to be conditionally $\sigma^2$-sub-Gaussian given another random variable $V$ (i.e., under $P_{U|V}$) for some $\sigma^2>0$ if we have almost surely that
    \begin{equation}\label{SGCC}
        \log\mathbb{E}[e^{\lambda(U-\mathbb{E}[U|V])}|V]\leq \frac{\sigma^2\lambda^2}{2}
    \end{equation}
    for all $\lambda\in\mathbb{R}$.
\end{definition}

Throughout the paper, we omit stating explicitly that the conditional sub-Gaussian inequality holds almost surely for the sake of simplicity.  

\section{Bounding Excess Minimum Risk}\label{sec:Main}

In this section, we establish a series of bounds on the excess minimum risk based on different information-theoretic measures. Our approach combines the variational characterizations of the KL divergence~\cite{DV}, the R\'enyi divergence~\cite{DVRformula}, and the Sibson mutual information~\cite[Theorem 2]{Esposito1}, along with the auxiliary distribution method introduced in~\cite{Aminian}.

\subsection{Renyi's Divergence Based Upper Bound}\label{sec-renyi-div-bnd}

\noindent We first  state the variational characterization of the R\'enyi divergence~\cite{DVRformula}, which generalizes the Donsker-Varadhan variational formula for KL divergence~\cite{DV}.

\smallskip

\begin{lemma}{\cite[Theorem 3.1]{DVRformula}}
    Let $P$ and $Q$ be two probability measures on $(\Omega,\mathcal{M})$ and $\alpha\in(0,\infty)$, $\alpha\neq 1$. Let $g$ be a measurable function such that $e^{(\alpha-1)g}\in\mathcal{L}^1(P)$ and $e^{\alpha g}\in\mathcal{L}^1(Q)$, where $\mathcal{L}^1(\mu)$ denotes the collection of all measurable functions with finite $\mathcal{L}^1$-norm. Then,
    \begin{equation}\label{RVRformula}
        D_{\alpha}(P\|Q) \geq \frac{\alpha}{\alpha-1}\log\mathbb{E}_{P}[e^{(\alpha-1)g(X)}] - \log\mathbb{E}_{Q}[e^{\alpha g(X)}].
    \end{equation}
\end{lemma}

\smallskip 

We next provide the following lemma, whose proof is a slight generalization of~\cite[Lemma 2]{Modak} and~\cite[Lemma~1]{Gyorfi}. 

\begin{lemma}\label{Decoup}
Consider two arbitrary jointly distributed random variables $U$ and $V$ defined on the same probability and taking values in spaces $\mathcal{U}$ and $\mathcal{V}$, respectively. Given a measurable function $h:\mathcal{U}\times \mathcal{V} \to \mathbb{R}$, assume that $h(u,V)$ is $\sigma^2(u)$-sub-Gaussian under $P_V$ and $P_{V|U=u}$ for all $u\in\mathcal{U}$, where $\mathbb{E}[\sigma^2(U)]<\infty$. Then for $\alpha\in(0,1)$,
    \begin{align*}
    |\mathbb{E}[{h(U,V)}]-&\mathbb{E}[{h(\bar{U},\bar{V})}]| \leq \sqrt{2\mathbb{E}[\sigma^2(U)] \, \frac{D_{\alpha}(P_{V|U}\|P_V|P_U)}{\alpha}},
    \end{align*}
    where $\bar{U}$ and $\bar{V}$ are independent copies of $U$ and $V$, respectively, (i.e. $P_{\bar{U},\bar{V}} = P_U P_V$).
\end{lemma}
\smallskip\noindent
{\bf Proof.}
    By the sub-Gaussian property, we have that
    \begin{equation}\label{SGL1}
        \log\mathbb{E}\left[e^{(\alpha-1)\lambda h(u,V)-\mathbb{E}[(\alpha-1)\lambda h(u,V)]|u=u}|U=u\right] \leq \frac{\lambda^2(\alpha-1)^2\sigma^2(u)}{2}
    \end{equation}
    and
    \begin{equation}\label{SGL}
        \log\mathbb{E}[e^{\alpha\lambda h(u,V)-\mathbb{E}[\alpha\lambda h(u,V)]}]\leq \frac{\lambda^2\alpha^2\sigma^2(u)}{2}.
    \end{equation}
    Re-arranging the terms gives us 
    \begin{equation}\label{SGLCond}
        -\log\mathbb{E}\left[e^{(\alpha-1)\lambda h(u,V)}|U=u\right] \geq -\frac{\lambda^2(\alpha-1)^2\sigma^2(u)}{2}+\mathbb{E}[(1-\alpha)\lambda h(u,V)|U=u]
    \end{equation}
    and
    \begin{equation}\label{RSGL}
        -\log\mathbb{E}[e^{\alpha\lambda h(u,V)}]\geq -\frac{\lambda^2\alpha^2\sigma^2(u)}{2}-\mathbb{E}[\alpha\lambda h(u,V)].
    \end{equation}
    Note that by \eqref{SGL1} and \eqref{SGL}, $e^{(\alpha-1)\lambda h(u,V)}\in\mathcal{L}^1(P_{V|U=u})$ and $e^{\alpha\lambda h(u,V)}\in\mathcal{L}^1(P_{V})$. 
    By the variational formula in~\eqref{RVRformula}, we have that
    \begin{align}
     D_{\alpha}(P_{V|U=u}\| P_V) \geq  \frac{\alpha}{\alpha-1}\log\mathbb{E}[e^{(\alpha-1)\lambda h(u,V)}|U=u]  - \log\mathbb{E}[e^{\alpha \lambda h(u,V)}]. \label{renyi-lb}
    \end{align}
    Substituting \eqref{SGLCond} and \eqref{RSGL} in~\eqref{renyi-lb} yields
    \begin{align*}
         D_{\alpha}(P_{V|U=u}\|P_V) &\geq  \frac{\alpha}{1-\alpha}\left(-\frac{\lambda^2(\alpha-1)^2\sigma^2(u)}{2}+\mathbb{E}[(1-\alpha)\lambda h(u,V)|U=u]\right) \\
        &\quad -\frac{\lambda^2\alpha^2\sigma^2(u)}{2}-\mathbb{E}[\alpha\lambda h(u,V)] \\
        &= \alpha\lambda(\mathbb{E}[h(u,V)|U=u]-\mathbb{E}[h(u,V)])
        -\frac{\lambda^2\alpha(1-\alpha)\sigma^2(u)}{2}\\ 
        &\mbox{} \quad -\frac{\lambda^2\alpha^2\sigma^2(u)}{2} \\
        &= \alpha\lambda(\mathbb{E}[h(u,V)|U=u]-\mathbb{E}[h(u,V)]) -\frac{\lambda^2\alpha\sigma^2(u)}{2} .         
    \end{align*}
The left-hand side of the resulting inequality 
\begin{align*}
    \frac{\lambda^2\alpha\sigma^2(u)}{2}-\alpha\lambda(\mathbb{E}[h(u,V)|U=u] 
    -\mathbb{E}[h(u,V)]) +D_{\alpha}(P_{V|U=u}\|P_V)\geq 0
\end{align*}
is a non-negative quadratic polynomial in  $\lambda$. Thus the discriminant is non-positive and we have 
\begin{align*}
    &(\alpha(\mathbb{E}[h(u,V)|U=u]-\mathbb{E}[h(u,V)]))^2 \leq 4\left(\frac{\alpha\sigma^2(u)}{2}\right)D_{\alpha}(P_{V|U=u}\|P_V).
\end{align*}
Therefore, 
\begin{align}\label{PLE}
    &|\mathbb{E}[h(u,V)|U=u]-\mathbb{E}[h(u,V)]| \leq \sqrt{\frac{2\sigma^2(u)D_{\alpha}(P_{V|U=u}\|P_V)}{\alpha}}.
\end{align}
Since, $\bar{U}$ and $\bar{V}$ are independent and $P_{\bar{V}}=P_V$, we have that
$$\mathbb{E}[h(u,V)] = \mathbb{E}[h(\bar{U},\bar{V})|\bar{U}=u].$$
Therefore, we have
\begin{align}
   |\mathbb{E}[h(U,V)] -\mathbb{E}[h(\bar{U},\bar{V})]|& = \left|\int(\mathbb{E}[h(U,V)|U=u]-\mathbb{E}[h(\bar{U},\bar{V})|\bar{U}=u])P_{U}(du)\right| \nonumber \\
   &= \left|\int(\mathbb{E}[h(u,V)|U=u]-\mathbb{E}[h(u,V)])P_{U}(du)\right| \nonumber\\
   &\leq \int\left|(\mathbb{E}[h(u,V)|U=u]-\mathbb{E}[h(u,V)]\right| P_{U}(du) \label{EJS1}\\
   &\leq \int \sqrt{\frac{2\sigma^2(u)D_{\alpha}(P_{V|U=u}\|P_V)}{\alpha}}P_{U}(du) \label{PLEU}\\
   &\leq \sqrt{\int2\sigma^2(u)P_{U}(du)}\sqrt{\int\frac{D_{\alpha}(P_{V|U=u}\|P_V)}{\alpha}P_{U}(du)} \label{CS1}\\
   &= \sqrt{2\mathbb{E}[\sigma^2(U)] \, \frac{D_{\alpha}(P_{V|U}\|P_V|P_U)}{\alpha}}, \label{LE}
\end{align}
where \eqref{EJS1} follows from Jensen's inequality, \eqref{PLEU} follows from \eqref{PLE}, \eqref{CS1} follows from the Cauchy-Schwarz inequality and the definition of conditional R\'enyi divergence  in~\eqref{CondRD} with $D_{\alpha}(P_{V|U}\|P_V|P_U) = \mathbb{E}_{U}[D_{\alpha}(P_{V|U}(\cdot|U)\|P_V)]$. 
\qed

\medskip

Note that the R\'enyi divergence based bound in Lemma~\ref{Decoup} differs from that in~\cite[Theorem 3]{Aminian}. In our approach, we consider sub-Gaussianity under both $P_V$ and  $P_{V|U=u}$ for all $u \in \mathcal{U}$, which allows for non-constant sub-Gaussian parameters. This leads to a more general bound that applies to a broader class of loss functions.

\smallskip

We next use Lemma~\ref{Decoup} to derive our theorem for the  R\'enyi divergence based bound; its proof is an adaptation of~\cite[Theorem 3]{Gyorfi}. 

\smallskip

\begin{theorem}\label{Mtheorem}
Let $X$, $Y$ and $Z$ be random vectors such that $Y\rightarrow X \rightarrow Z$, as described in Section~\ref{subsec:problem_setup}.  Assume that there exists an optimal estimator $f$ of Y from X such that $l(y,f(X))$ is conditionally $\sigma^2(y)$-sub-Gaussian under $P_{X|Z}$ and $P_{X|Z,Y=y}$ for all $y\in\mathbb{R}^p$, i.e.,
    $$\log\mathbb{E}[e^{(\lambda(l(y,f(X)))-\mathbb{E}[l(y,f(X))|Z])}|Z]\leq \frac{\sigma^2(y)\lambda^2}{2}$$ and $$\log\mathbb{E}[e^{(\lambda(l(y,f(X)))-\mathbb{E}[l(y,f(X))|Z,Y=y])}|Z,Y=y]\leq \frac{\sigma^2(y)\lambda^2}{2}$$ for all $\lambda\in\mathbb{R}$ and $y\in\mathbb{R}$, where $\sigma^2:\mathbb{R}\rightarrow\mathbb{R}$, satisfies $\mathbb{E}[\sigma^2(Y)]<\infty$. Then for $\alpha\in(0,1)$, the excess minimum risk satisfies
    \begin{align}\label{LT1}
        &L_{l}^{*}(Y|Z)-L_{l}^{*}(Y|X) \leq\sqrt{\frac{2\mathbb{E}[\sigma^2(Y)]}{\alpha} \, D_{\alpha}(P_{X|Y,Z} \|P_{X|Z}|P_{Y,Z})}.
    \end{align}
\end{theorem}

\smallskip\noindent
{\bf Proof.}
Let $\bar{X}$, $\bar{Y}$ and $\bar{Z}$ be random variables such that $P_{\bar{Y}|\bar{Z}} = P_{Y|Z}$, $P_{\bar{X}|\bar{Z}} = P_{X|Z}$, $P_{\bar{Z}} = P_{Z}$ and $\bar{Y}$ and $\bar{X}$ are conditionally independent given $\bar{Z}$, i.e., \\ $P_{\bar{Y},\bar{X},\bar{Z}} = P_{Y|Z}P_{X|Z}P_{Z}$. 

\smallskip

    We apply Lemma~\ref{Decoup} by setting $U=Y$, $V=X$ and $h(u,v)=l(y,f(x))$. Consider $\mathbb{E}[l(Y,f(X))|Z=z])$ and $\mathbb{E}[l(\bar{Y},f(\bar{X}))|Z=z]$ as regular expectations taken with respect to $P_{Y,X|Z=z}$ and $P_{\bar{Y},\bar{X}|Z=z}$. Since, $\bar{Y}$ and $\bar{X}$ are conditionally independent given $\bar{Z}=z$ and $P_{\bar{Z}} = P_{Z}$, we have that
    \begin{align}\label{lemma}
        \MoveEqLeft[8]|\mathbb{E}[l(Y,f'(X))|Z=z])-\mathbb{E}[l(\bar{Y},f(\bar{X}))|Z=z])|& \nonumber\\
        &\hspace{2cm}\leq\sqrt{\frac{2\mathbb{E}[\sigma^2(Y)|Z=z]}{\alpha}D_{\alpha}(P_{X|Y,Z=z}\|P_{X|Z=z}|P_{Y|Z=z})}.
    \end{align}  
    Now,
    \begin{align}
        \MoveEqLeft[10]\left|\mathbb{E}[l(Y,f(X))]-\mathbb{E}[l(\bar{Y},f(\bar{X}))]\right| &\nonumber \\
        &\leq \int\left|\mathbb{E}[l(Y,f(X))|Z=z])-\mathbb{E}[l(\bar{Y},f(\bar{X}))|Z=z])\right|P_{Z}(dz) \nonumber \\
        &\leq \int\Bigg( \sqrt{\frac{2\mathbb{E}[\sigma^2(Y)|Z=z]}{\alpha}} \nonumber \\
        &\hspace{1.5cm}\times \sqrt{D_{\alpha}(P_{X|Y,Z=z}\|P_{X|Z=z}|P_{Y|Z=z})}\, \Bigg) P_{Z}(dz) \nonumber \\
        &\leq\sqrt{2\int\mathbb{E}[\sigma^2(Y)|Z=z]P_{Z}(dz)}\nonumber\\
        &\hspace{1.5cm}\times\sqrt{\int\frac{D_{\alpha}(P_{X|Y,Z=z}\|P_{X|Z=z}|P_{Y|Z=z})}{\alpha}P_{Z}(dz)} \nonumber 
        \\
        &=\sqrt{\frac{2\mathbb{E}[\sigma^2(Y)]}{\alpha} \, D_{\alpha}(P_{X|Y,Z}\|P_{X|Z}|P_{Y,Z})} ,\label{TheoremR}
    \end{align}
    where the first inequality  follows from Jensen's inequality and since $P_{\bar{Z}} = P_{Z}$, the second inequality follows from~\eqref{lemma}, the third  from the Cauchy-Schwarz inequality,  and the equality follows from~\eqref{CondRD}. Since, $\bar{Y}$ and $\bar{X}$ are conditionally independent given $\bar{Z}$, we get the Markov chain $\bar{Y}\rightarrow \bar{Z} \rightarrow \bar{X}$. Then we have
    \begin{align}
        \mathbb{E}[l(\bar{Y},f(\bar{X}))])&\geq L_{l}^{*}(\bar{Y}|\bar{X}) \nonumber \\
        &\geq L_{l}^{*}(\bar{Y}|\bar{Z}) \nonumber \\
        &=L_{l}^{*}(Y|Z), \label{TI1}
    \end{align}
    where the first inequality follows since $\bar{Y}\rightarrow\bar{X}\rightarrow f(\bar{X})$, the second inequality holds since $\bar{Y}\rightarrow \bar{Z} \rightarrow \bar{X}$ by construction, and the equality follows since $(\bar{Y},\bar{Z})$ and $(Y,Z)$ have the same distribution by construction.
    Since, $f$ is an optimal estimator of $Y$ from $X$, we also have 
    \begin{equation}\label{TI2}
        \mathbb{E}[l(Y,f(X))])= L_{l}^{*}(Y|X).
    \end{equation}
    Therefore using~\eqref{TI1} and~\eqref{TI2} in~\eqref{TheoremR} combined with the fact that $L_{l}^{*}(Y|Z)\geq L_{l}^{*}(Y|X)$, we arrive at the desired inequality:
    \begin{align*}
        L_{l}^{*}(Y|Z)-L_{l}^{*}(Y|X) \leq\sqrt{\frac{2\mathbb{E}[\sigma^2(Y)]}{\alpha} \, D_{\alpha}(P_{X|Y,Z}\|P_{X|Z}|P_{Y,Z})}.
    \end{align*}
\qed

\medskip

\begin{remark}\label{RemMIB}
    Taking the limit as $\alpha \rightarrow 1$ of the right-hand side of~\eqref{LT1} in Theorem~\ref{Mtheorem}, we have that 
    \begin{align}
        L_{l}^{*}(Y|Z)-L_{l}^{*}(Y|X)
        &\leq\sqrt{2\mathbb{E}[\sigma^2(Y)] \, D_{\text{KL}}(P_{X|Y,Z}\|P_{X|Z}|P_{Y,Z})} \nonumber\\
        &= \sqrt{2\mathbb{E}[\sigma^2(Y)] \, (I(X;Y)-I(Z;Y))},\label{MIB}
    \end{align}
   recovering the bound in~\cite[Theorem 3]{Gyorfi}.
\end{remark}

\smallskip

As a special case, we consider bounded loss functions, which naturally satisfy the conditional sub-Gaussian condition. The following corollary is an application of Theorem~\ref{Mtheorem} under a fixed sub-Gaussian parameter. For completeness, we include the full proof.

\smallskip

\begin{corollary}\label{CorR}
    Suppose the loss function $l$ is bounded, i.e., $\|l\|_{\infty} =\sup_{y,y'} l(y,y')< \infty$. Then for random vectors $X$, $Y$ and $Z$ such that $Y\rightarrow X \rightarrow Z$ as described in Section~\ref{subsec:problem_setup}, we have the following inequality for $\alpha\in(0,1)$ on the excess minimum risk:
    \begin{equation}\label{CR1}
        L_{l}^{*}(Y|Z)-L_{l}^{*}(Y|X)\leq \frac{\|l\|_{\infty}}{\sqrt{2}}\sqrt{\frac{D_{\alpha}(P_{X|Y,Z}\|P_{X|Z}|P_{Y,Z})}{\alpha}}.
    \end{equation}
\end{corollary}
\smallskip\noindent
{\bf Proof.}
We show that the bounded loss function $l$ satisfies the conditional sub-Gaussian properties of Theorem~\ref{Mtheorem}. Since $l$ is bounded we have that for any $f:\mathbb{R}^q\rightarrow\mathbb{R}^p$, $x\in\mathbb{R}^q$ and $y\in\mathbb{R}^p$, $l(y,f(x))\in[0,\|l\|_{\infty}]$. Then by Hoeffding's lemma~\cite{Hoeffding}, we can write 
 $$\log\mathbb{E}[e^{(\lambda(l(y,f(X)))}|Z]\leq \mathbb{E}[\lambda l(y,f(X))|Z])+ \frac{\|l\|_{\infty}^2\lambda^2}{8}$$ and 
 \begin{align*}
     &\log\mathbb{E}[e^{(\lambda(l(y,f(X)))}|Z,Y=y] \leq \mathbb{E}[\lambda l(y,f(X))|Z,Y=y])+ \frac{\|l\|_{\infty}^2\lambda^2}{8}
 \end{align*}
 for all $\lambda\in\mathbb{R}$ and $y\in\mathbb{R}$. Rearranging the above inequalities gives us that $l(y,f(X))$ is conditionally $\frac{\|l\|_{\infty}^2}{4}$-sub-Gaussian under both $P_{X|Z}$ and $P_{X|Z,Y=y}$ for all $y\in\mathbb{R}^p$. Then by~\eqref{LT1}, we have
\begin{equation}
        L_{l}^{*}(Y|Z)-L_{l}^{*}(Y|X)\leq \frac{\|l\|_{\infty}}{\sqrt{2}}\sqrt{\frac{D_{\alpha}(P_{X|Y,Z}\|P_{X|Z}|P_{Y,Z})}{\alpha}}. \label{bnd-renyi}
    \end{equation}
\qed

\medskip

\begin{remark}
    Taking the limit as $\alpha \rightarrow 1$ of~\eqref{bnd-renyi} in Corollary~\ref{CorR} yields the mutual information based bound:
    \begin{align}
        L_{l}^{*}(Y|Z)-L_{l}^{*}(Y|X)&\leq \frac{\|l\|_{\infty}}{\sqrt{2}}\sqrt{D_{\text{KL}}(P_{X|Y,Z}\|P_{X|Z}|P_{Y,Z})} \nonumber\\
        &= \frac{\|l\|_{\infty}}{\sqrt{2}}\sqrt{I(X;Y)-I(Z;Y)},\label{MIB1}
    \end{align}
    which recovers the bound in~\cite[Corollary~1]{Gyorfi}.
\end{remark}

\subsection{$\bm{\alpha}$-Jensen-Shannon Divergence Based Upper Bound}

\smallskip

We next derive $\alpha$- Jensen-Shannon divergence based bounds on minimum excess risk.  

\smallskip

We consider two arbitrary jointly distributed random variables $U$ and $V$ defined on the same probability space and taking values in $\mathcal{U}$ and $\mathcal{V}$, respectively. Throughout this section, we work with the joint distribution $P_{U,V}$ over $\mathcal{U} \times \mathcal{V}$ and the corresponding product of marginals $P_{U}P_{V}$. For convenience, we also define additional distributions that will play an important role in the derivation of our bounds.

 \begin{definition}\label{AJScondist}
 The $\alpha$-convex combination of the joint distribution $P_{U,V}$
and the product of two marginals $P_{U} P_{V}$ is denoted by $P^{(\alpha)}_{U,V}$ and given by
\begin{equation}
    P^{(\alpha)}_{U,V} = \alpha P_{U,V}+(1-\alpha)P_{U} P_{V}
\end{equation}
for $\alpha\in(0,1)$.
\end{definition}

\smallskip

\begin{definition}
    The $\alpha$-conditional convex combination of the conditional distribution $P_{V|U}$
and the marginal $P_{V}$ is denoted by $P^{(\alpha)}_{V|U}$ and given by
\begin{equation}\label{Acondcondist}
    P^{(\alpha)}_{V|U} = \alpha P_{V|U}+(1-\alpha)P_{V}
\end{equation}
for $\alpha\in(0,1)$.
\end{definition}

\smallskip

\noindent We first provide the following lemma, whose proof is a slight generalization of~\cite[Lemma 2]{Aminian} and \cite[Lemma~1]{Gyorfi}.

\begin{lemma}\label{DecoupJS}
 Given a function $h:\mathcal{U}\times \mathcal{V} \to \mathbb{R}$, assume that $h(u,V)$ is $\sigma^2(u)$-sub-Gaussian under $P^{(\alpha)}_{V|U=u}$ for all $u\in\mathcal{U}$, where $\mathbb{E}[\sigma^2(U)]<\infty$. Then for $\alpha\in(0,1)$,
    \begin{align*}
    |\mathbb{E}_{P_{U,V}}[{h(U,V)}]-&\mathbb{E}_{P_{U}P_{V}}[{h(U,V)}]| 
    \leq \sqrt{2\mathbb{E}[\sigma^2(U)] \, \frac{JS_{\alpha}(P_{U,V}\|P_UP_V)}{\alpha(1-\alpha)}}.
    \end{align*}
\end{lemma}
\smallskip\noindent
{\bf Proof.}
    Using the assumption that the function $h(u,V)$ is $\sigma^2(u)$-sub-Gaussian under $P^{(\alpha)}_{V|U=u}$ for all $u\in\mathcal{U}$ and the  Donsker-Varadhan representation~\cite{DV} for $D_{\text{KL}}(P_{V|U=u}\|P^{(\alpha)}_{V|U=u})$, we have that
    \begin{eqnarray}
    D_{\text{KL}}(P_{V|U=u}\|P^{(\alpha)}_{V|U=u}) &\geq& \mathbb{E}_{P_{V|U=u}}[\lambda h(u,V)] - \log\mathbb{E}_{P^{(\alpha)}_{V|U=u}}[e^{\lambda h(u,V)}] \nonumber \\
    &\geq& \mathbb{E}_{P_{V|U=u}}[\lambda h(u,V)] - \mathbb{E}_{P^{(\alpha)}_{V|U=u}}[\lambda h(u,V)] - \frac{\lambda^2\sigma^2(u)}{2},
\end{eqnarray}
for all $\lambda\in\mathbb{R}$. Rearranging terms, we obtain for all $\lambda\in\mathbb{R}$ and $u\in\mathcal{U}$ that 
    \begin{equation*}
        \lambda\left(\mathbb{E}_{P_{V|U=u}}[h(u,V)] - \mathbb{E}_{P^{(\alpha)}_{V|U=u}}[h(u,V)]\right) \leq D_{\text{KL}}(P_{V|U=u}\|P^{(\alpha)}_{V|U=u})+\frac{\lambda^2\sigma^2(u)}{2}.
    \end{equation*}
Similarly, using the assumption that the function $h(u,V)$ is $\sigma^2(u)$-sub-Gaussian under $P^{(\alpha)}_{V|U=u}$ for all $u\in\mathcal{U}$ and the Donsker-Varadhan representation~\cite{DV} for $D_{\text{KL}}(P_{V}\|P^{(\alpha)}_{V|U=u})$, we have for all $\lambda'\in\mathbb{R}$ that
    \begin{equation*}
        \lambda'\left(\mathbb{E}_{P_{V}}[h(u,V)] - \mathbb{E}_{P^{(\alpha)}_{V|U=u}}[h(u,V)]\right) \leq D_{\text{KL}}(P_{V}\|P^{(\alpha)}_{V|U=u})+\frac{\lambda'^2\sigma^2(u)}{2}.
    \end{equation*}
If $\lambda<0$, then taking $\lambda'=\frac{\alpha}{\alpha-1}\lambda>0$ yields that   
\begin{equation}\label{Pr1e1}
    \mathbb{E}_{P^{(\alpha)}_{V|U=u}}[h(u,V)] - \mathbb{E}_{P_{V|U=u}}[h(u,V)] \leq \frac{D_{\text{KL}}(P_{V|U=u}\|P^{(\alpha)}_{V|U=u})}{|\lambda|}+\frac{|\lambda|\sigma^2(u)}{2}, 
\end{equation}
and 
\begin{equation}\label{Pr1e2}
    \mathbb{E}_{P_{V}}[h(u,V)] - \mathbb{E}_{P^{(\alpha)}_{V|U=u}}[h(u,V)]  \leq \frac{D_{\text{KL}}(P_{V}\|P^{(\alpha)}_{V|U=u})}{\lambda'}+\frac{\lambda'\sigma^2(u)}{2}.
\end{equation}
Adding \eqref{Pr1e1} and \eqref{Pr1e2} yields for all $\lambda<0$ that
\begin{align}\label{e1}
  \MoveEqLeft[6]\mathbb{E}_{P_{V}}[h(u,V)] - \mathbb{E}_{P_{V|U=u}}[h(u,V)]&    \nonumber \\
  & \leq \frac{D_{\text{KL}}(P_{V|U=u}\|P^{(\alpha)}_{V|U=u})}{|\lambda|}+\frac{|\lambda|\sigma^2(u)}{2} + \frac{D_{\text{KL}}(P_{V}\|P^{(\alpha)}_{V|U=u})}{\lambda'}+\frac{\lambda'\sigma^2(u)}{2},  \nonumber \\
  &= \frac{D_{\text{KL}}(P_{V|U=u}\|P^{(\alpha)}_{V|U=u})}{|\lambda|}+\frac{|\lambda|\sigma^2(u)}{2} + \frac{D_{\text{KL}}(P_{V}\|P^{(\alpha)}_{V|U=u})}{\frac{\alpha}{\alpha-1}\lambda}+\frac{\frac{\alpha}{\alpha-1}\lambda\sigma^2(u)}{2}, \nonumber \\
  &= \frac{\alpha D_{\text{KL}}(P_{V|U=u}\|P^{(\alpha)}_{V|U=u})+(1-\alpha)D_{\text{KL}}(P_{V}\|P^{(\alpha)}_{V|U=u})}{\alpha |\lambda|}+\frac{|\lambda|\sigma^2(u)}{2(1-\alpha)}. 
\end{align}

\smallskip

\noindent Similarly for $\lambda>0$, taking $\lambda'=\frac{\alpha}{\alpha-1}\lambda<0$, we have
\begin{equation}\label{-Pr1e1}
    -\left(\mathbb{E}_{P^{(\alpha)}_{V|U=u}}[h(u,V)] - \mathbb{E}_{P_{V|U=u}}[h(u,V)]\right) \leq \frac{D_{\text{KL}}(P_{V|U=u}\|P^{(\alpha)}_{V|U=u})}{\lambda}+\frac{\lambda\sigma^2(u)}{2}, 
\end{equation}
and 
\begin{equation}\label{-Pr1e2}
    -\left(\mathbb{E}_{P_{V}}[h(u,V)] - \mathbb{E}_{P^{(\alpha)}_{V|U=u}}[h(u,V)]\right)  \leq \frac{D_{\text{KL}}(P_{V}\|P^{(\alpha)}_{V|U=u})}{|\lambda'|}+\frac{|\lambda'|\sigma^2(u)}{2}.
\end{equation}
\smallskip
Adding \eqref{-Pr1e1} and \eqref{-Pr1e2} yields for all $\lambda>0$ that
\begin{align}\label{Pr1l2}
    \MoveEqLeft[6]-\left( \mathbb{E}_{P_{V}}[h(u,V)] - \mathbb{E}_{P_{V|U=u}}[h(u,V)]\right)& \nonumber \\
    & \leq \frac{\alpha D_{\text{KL}}(P_{V|U=u}\|P^{(\alpha)}_{V|U=u})+(1-\alpha)D_{\text{KL}}(P_{V}\|P^{(\alpha)}_{V|U=u})}{\alpha \lambda}+\frac{\lambda\sigma^2(u)}{2(1-\alpha)} .
\end{align}
 From \eqref{e1} and \eqref{Pr1l2}, we get the following non-negative parabola in terms of $\lambda$:
\begin{align}\label{NGPJS}
   \MoveEqLeft[8]   \lambda^2\left(\frac{\sigma^2(u)}{2(1-\alpha)}\right) +\lambda\left( \mathbb{E}_{P_{V}}[h(u,V)] - \mathbb{E}_{P_{V|U=u}}[h(u,V)]\right) \nonumber \\ &+ \frac{\alpha D_{\text{KL}}(P_{V|U=u}\|P^{(\alpha)}_{V|U=u})+(1-\alpha)D_{\text{KL}}(P_{V}\|P^{(\alpha)}_{V|U=u})}{\alpha} \geq 0,
\end{align}
\noindent where~\eqref{NGPJS} holds trivially for $\lambda=0$. Thus its discriminant is non-positive, and we have for all $\alpha\in(0,1)$ that
\begin{align*}
    \MoveEqLeft[6]|\mathbb{E}_{P_{V}}[h(u,V)] - \mathbb{E}_{P_{V|U=u}}[h(u,V)]|^2&  \\
    &\leq 2\sigma^2(u)\frac{\left(\alpha D_{\text{KL}}(P_{V|U=u}\|P^{(\alpha)}_{V|U=u})+(1-\alpha)D_{\text{KL}}(P_{V}\|P^{(\alpha)}_{V|U=u})\right)}{\alpha(1-\alpha)}.
\end{align*}
Hence,
\begin{align}\label{PLEAJS}
    \MoveEqLeft[6]|\mathbb{E}_{P_{V}}[h(u,V)] - \mathbb{E}_{P_{V|U=u}}[h(u,V)]| & \nonumber\\*
    &\leq \sqrt{2\sigma^2(u)\frac{\left(\alpha D_{\text{KL}}(P_{V|U=u}\|P^{(\alpha)}_{V|U=u})+(1-\alpha)D_{\text{KL}}(P_{V}\|P^{(\alpha)}_{V|U=u})\right)}{\alpha(1-\alpha)}}.
\end{align}
As a result, we obtain that
\begin{align}
    \MoveEqLeft[1]|\mathbb{E}_{P_{U,V}}[{h(U,V)}]-\mathbb{E}_{P_{U}P_{V}}[{h(U,V)}]| \nonumber \\
    &= \left|\int(\mathbb{E}_{P_{V}}[h(u,V)] - \mathbb{E}_{P_{V|U=u}}[h(u,V)])P_{U}(du)\right| \nonumber \\
    &\leq \int\left|\mathbb{E}_{P_{V}}[h(u,V)] - \mathbb{E}_{P_{V|U=u}}[h(u,V)]\right|P_{U}(du) \label{JS1}\\
    &\leq \int\left(\sqrt{2\sigma^2(u)\frac{\left(\alpha D_{\text{KL}}(P_{V|U=u}\|P^{(\alpha)}_{V|U=u})+(1-\alpha)D_{\text{KL}}(P_{V}\|P^{(\alpha)}_{V|U=u})\right)}{\alpha(1-\alpha)}}\right)P_{U}(du) \label{PLEU1}\\
    &\leq \sqrt{\int2\sigma^2(u)P_{U}(du)} \nonumber \\
    &\hspace{0.5cm} \times \sqrt{\int\frac{\left(\alpha D_{\text{KL}}(P_{V|U=u}\|P^{(\alpha)}_{V|U=u})+(1-\alpha)D_{\text{KL}}(P_{V}\|P^{(\alpha)}_{V|U=u})\right)}{\alpha(1-\alpha)}P_{U}(du)} \label{CS1JS}\\
    &= \sqrt{\mathbb{E}[2\sigma^2(U)]\frac{\left(\alpha D_{\text{KL}}(P_{U,V}\|P^{(\alpha)}_{U,V})+(1-\alpha)D_{\text{KL}}(P_{U}P_{V}\|P^{(\alpha)}_{U,V})\right)}{\alpha(1-\alpha)}} \label{CS2}\\
    &= \sqrt{\mathbb{E}[2\sigma^2(U)] \, \frac{JS_{\alpha}(P_{U,V}\|P_UP_V)}{\alpha(1-\alpha)}}, \label{LEJS}
\end{align}
where \eqref{JS1} follows from Jensen's inequality, \eqref{PLEU1} follows from \eqref{PLEAJS}, \eqref{CS1JS} follows from the Cauchy-Schwarz inequality, \eqref{CS2} follows from the definition of the conditional KL divergence and Definition \ref{AJScondist} and \eqref{LEJS} follows from the definition of $\alpha$-Jensen-Shannon divergence.
\qed

\medskip

\noindent We next use Lemma~\ref{DecoupJS} to derive our  theorem for $\alpha$-Jensen-Shannon divergence- based bound.

\smallskip

\begin{theorem}\label{MtheoremJS}
Let $X$, $Y$ and $Z$ be random vectors such that $Y\rightarrow X \rightarrow Z$, as described in Section~\ref{subsec:problem_setup}.  Assume that there exists an optimal estimator $f$ of Y from X such that $l(y,f(X))$ is conditionally $\sigma^2(y)$-sub-Gaussian under $P^{(\alpha)}_{X|Z,Y=y} = \alpha P_{X|Z,Y=y}+(1-\alpha)P_{X|Z}$ for all $y\in\mathbb{R}^p$, i.e.,
$$\log\mathbb{E}_{P^{(\alpha)}_{X|Z,Y=y}}\left[e^{\lambda\big( l(y,f(X))-\mathbb{E}_{P^{(\alpha)}_{X|Z,Y=y}}\big[ l(y,f(X))\big]\big)}\right]  \leq \frac{\sigma^2(y)\lambda^2}{2}$$ for all $\lambda\in\mathbb{R}$ and $y\in\mathbb{R}$, where $\sigma^2:\mathbb{R}\rightarrow\mathbb{R}$ satisfies $\mathbb{E}[\sigma^2(Y)]<\infty$. Then for $\alpha\in(0,1)$, the excess minimum risk satisfies
    \begin{align}\label{LT1JS}
        &L_{l}^{*}(Y|Z)-L_{l}^{*}(Y|X) 
\leq\sqrt{\frac{2\mathbb{E}[\sigma^2(Y)]}{\alpha(1-\alpha)} \, JS_{\alpha}(P_{Y,X|Z} \|P_{Y|Z}P_{X|Z}|P_{Z})}.
    \end{align}
\end{theorem}
\smallskip\noindent
{\bf Proof.}
    Let $\bar{X}$, $\bar{Y}$ and $\bar{Z}$ be the same random variables defined in Theorem~\ref{Mtheorem}.

    \noindent Similar to the proof of Theorem~\ref{Mtheorem}, we apply Lemma~\ref{DecoupJS} by setting $U=Y$, $V=X$ and $h(u,v)=l(y,f(x))$ and taking regular expectations with respect to $P_{Y,X|Z=z}$ and $P_{\bar{Y},\bar{X}|Z=z}$. Since, $\bar{Y}$ and $\bar{X}$ are conditionally independent given $\bar{Z}=z$ such that $P_{\bar{Y},\bar{X}|Z=z} = P_{Y|Z}P_{X|Z}$ and $P_{\bar{Z}} = P_{Z}$, we have that
    \begin{align}\label{lemmaJS}
        \MoveEqLeft[8]|\mathbb{E}_{P_{Y,X|Z=z}}[l(Y,f(X))]-\mathbb{E}_{P_{\bar{Y},\bar{X}|Z=z}}[l(\bar{Y},f(\bar{X}))]|& \nonumber \\ &= |\mathbb{E}_{P_{Y,X|Z=z}}[l(Y,f(X))]-\mathbb{E}_{P_{Y|Z=z}P_{X|Z=z}}[l(Y,f(X))]| \nonumber\\
        &\leq\sqrt{\frac{2\mathbb{E}[\sigma^2(Y)|Z=z]}{\alpha(1-\alpha)}JS_{\alpha}(P_{Y,X|Z=z}\|P_{X|Z=z}P_{Y|Z=z})}.
    \end{align}  
    Thus,
    \begin{align}
        \MoveEqLeft[3]\left|\mathbb{E}_{P_{Y,X}}[l(Y,f(X))]-\mathbb{E}_{P_{\bar{Y},\bar{X}}}[l(\bar{Y},f(\bar{X}))]\right| & \nonumber \\
        & \leq \int\left|\mathbb{E}_{P_{Y,X|Z=z}}[l(Y,f(X))]-\mathbb{E}_{P_{\bar{Y},\bar{X}|Z=z}}[l(\bar{Y},f(\bar{X}))])\right|P_{Z}(dz) \nonumber \\
        &\leq \int\Bigg( \sqrt{\frac{2\mathbb{E}[\sigma^2(Y)|Z=z]}{\alpha(1-\alpha)}JS_{\alpha}(P_{Y,X|Z=z}\|P_{X|Z=z}P_{Y|Z=z})}\, \Bigg) P_{Z}(dz) \nonumber \\
        &\leq\sqrt{2\int\mathbb{E}[\sigma^2(Y)|Z=z]P_{Z}(dz)}\sqrt{\int\frac{JS_{\alpha}(P_{Y,X|Z=z}\|P_{X|Z=z}P_{Y|Z=z})}{\alpha(1-\alpha)}P_{Z}(dz)} \nonumber 
        \\
        &=\sqrt{\frac{2\mathbb{E}[\sigma^2(Y)]}{\alpha(1-\alpha)} \, JS_{\alpha}(P_{Y,X|Z} \|P_{Y|Z}P_{X|Z}|P_{Z})} ,\label{TheoremJS}
    \end{align}
    where the first inequality  follows from Jensen's inequality and since $P_{\bar{Z}} = P_{Z}$, the second inequality follows from~\eqref{lemmaJS}, the third  from the Cauchy-Schwarz inequality,  and the equality follows from~\eqref{AJSCond}.
    From proof of Theorem~\ref{Mtheorem}, we know that $\bar{Y}\rightarrow \bar{Z} \rightarrow \bar{X}$ forms a Markov chain. Hence,
    \begin{align}
        \mathbb{E}_{P_{\bar{Y},\bar{X}}}[l(\bar{Y},f(\bar{X}))])\geq L_{l}^{*}(Y|Z), \label{TI1JS}
    \end{align}
    Since, $f$ is an optimal estimator of $Y$ from $X$, we also have
    \begin{equation}\label{TI2JS}
        \mathbb{E}_{P_{Y,X}}[l(Y,f(X))])= L_{l}^{*}(Y|X).
    \end{equation}
    Therefore using~\eqref{TI1JS} and~\eqref{TI2JS} in~\eqref{TheoremJS} combined with the fact that $L_{l}^{*}(Y|Z)\geq L_{l}^{*}(Y|X)$, we arrive at the desired inequality:
    \begin{align*}
        &L_{l}^{*}(Y|Z)-L_{l}^{*}(Y|X) \leq\sqrt{\frac{2\mathbb{E}[\sigma^2(Y)]}{\alpha(1-\alpha)} \, JS_{\alpha}(P_{Y,X|Z} \|P_{Y|Z}P_{X|Z}|P_{Z})}.
    \end{align*}
\qed

\begin{remark}\label{RemAJS1}
Taking the limit as $\alpha \rightarrow 0$ on the right-hand side of~\eqref{LT1JS} in Theorem~\ref{MtheoremJS},
we obtain
\begin{align*}
    L_{l}^{*}(Y| Z) - L_{l}^{*}(Y | X)
    &\le \sqrt{2 \, \mathbb{E}[\sigma^2(Y)] \, (I(X;Y) - I(Z;Y))}, \label{MIBCJS}
\end{align*}
recovering the bound~\eqref{MIB} of~\cite[Theorem 3]{Gyorfi}. Furthermore, taking the limit as $\alpha \rightarrow 1$ of the right-hand side of~\eqref{LT1JS} in Theorem~\ref{MtheoremJS}, yields 
    \begin{align}
        L_{l}^{*}(Y|Z)-L_{l}^{*}(Y|X) 
        &\leq\sqrt{2\mathbb{E}[\sigma^2(Y)] \, D_{\text{KL}}(P_{Y|Z}P_{X|Z} \|P_{Y,X|Z}|P_{Z})} \nonumber\\
        &= \sqrt{2\mathbb{E}[\sigma^2(Y)] \, (L(X;Y)-L(Z;Y))},
    \end{align}
    where $L(U;V) := D_{\text{KL}}(P_U P_V \| P_{U,V})$ is called the Lautum information between $U$ and $V$, defined as the reverse KL divergence (i.e., the KL divergence between the product of marginals and the joint distribution)~\cite{palomar2008lautum}. We therefore obtain an upper bound on the minimum excess risk in terms of the reverse KL divergence.

\end{remark}

\smallskip

We close this section by specializing Theorem~\ref{MtheoremJS} to the case of bounded loss functions, hence obtaining a counterpart result to Corollary~\ref{CorR}.

\smallskip

\begin{corollary}\label{CorJS}
    Suppose the loss function $l$ is bounded. Then for random vectors $X$, $Y$ and $Z$ such that $Y\rightarrow X \rightarrow Z$ as described in Section~\ref{subsec:problem_setup}, we have the following inequality for $\alpha\in(0,1)$ on the excess minimum risk:
    \begin{equation}\label{CR1JS}
        L_{l}^{*}(Y|Z)-L_{l}^{*}(Y|X)\leq \frac{\|l\|_{\infty}}{\sqrt{2}}\sqrt{\frac{JS_{\alpha}(P_{Y,X|Z} \|P_{Y|Z}P_{X|Z}|P_{Z})}{\alpha(1-\alpha)}}.
    \end{equation}
\end{corollary}

\smallskip

\subsection{Sibson Mutual Information Based Upper Bound}

\smallskip
We herein bound the excess minimum risk based on Sibson's mutual information. We recall from Definition~\ref{ASibM2} that $U$ and $V$ are jointly distributed on measurable spaces $\mathcal{U}$ and $\mathcal{V}$, with joint distribution $P_{U,V}$ and marginals $P_U$ and $P_V$, assuming that all distributions are absolutely continuous with respect to a common sigma-finite measure $\nu$, with densities $p_U$, $p_V$, and $p_{U,V}$. Let $U^*$ denote the random variable whose distribution $P_{U^*}$ attains the minimum in the definition of the Sibson mutual information $I_\alpha^S(P_{U,V}, P_{U^*})$, with density $p_{U^*}$ as given in~\eqref{ASibMd2prime}. We now define an auxiliary distribution that will be central to the derivation of the main bounds in this section.

\smallskip

\begin{definition}\label{hatpdf}
Let $P_{\hat{U},\hat{V}}$ be a joint distribution on  $\mathcal{U}\times\mathcal{V}$ determined by density $p_{\hat{U},\hat{V}}$ that is obtained by tilting (using parameter $\alpha$) the densities $p_{U,V}$, $p_{U^*}$ and $p_{V}$ as follows:
\begin{equation}
    p_{\hat{U},\hat{V}}(u,v) = \frac{(p_{U,V}(u,v))^{\alpha}(p_{U^*}(u)p_{V}(v))^{(1-\alpha)}}{\iint(p_{U,V}(u',v'))^{\alpha}(p_{U^*}(u')p_{V}(v'))^{(1-\alpha)}du'dv'}
\end{equation}
for $\alpha\in(0,1)$.
\end{definition}

\smallskip

\noindent We state the following lemma based on the variational representation of the Sibson mutual information~\cite[Theorem 2]{Esposito1}, which establishes a connection to the KL divergence. The proof of the lemma follows from~\cite[Lemma 3]{Aminian} and~\cite[Theorem 5.1]{Esposito2}.

\begin{lemma}\label{Asiblem}
For the distributions $P_{\hat{U},\hat{V}}$, $P_{U,V}$ and $P_{U^*}P_{V}$ we have
    $$\alpha D_{\text{KL}}(P_{\hat{U},\hat{V}}\|P_{U,V})+(1-\alpha)D_{\text{KL}}(P_{\hat{U},\hat{V}}\|P_{U^*}P_{V}) = (1-\alpha)I_{\alpha}^{S}(P_{U,V},P_{U^*}).$$ 
\end{lemma}

\smallskip

\noindent We now invoke a basic but important property of sub-Gaussian random variables that will be used later in our analysis. Specifically, the set of all sub-Gaussian random variables has a linear structure. This property is well established in the literature~\cite{buldygin1980sub,rivasplata2012subgaussian}.

\begin{lemma}\label{SubAdd}
If $X$ is $\sigma_X^2$-sub-Gaussian random variable, then for any $\alpha\in\mathbb{R}$, the random variable
$\alpha X$ is $|\alpha|\sigma_X^2$-sub-Gaussian. If $Y$ is $\sigma_Y^2$-sub-Gaussian random variable, then the sum $X+Y$ is sub-Gaussian with parameter $(\sigma_X + \sigma_Y)^2$.
\end{lemma}

\smallskip

\noindent We next provide the following lemma, whose proof is a slight generalization of~\cite[Theorem 4]{Aminian} and ~\cite[Lemma 1]{Gyorfi}

\begin{lemma}\label{DecoupAS}
 Given a function $h:\mathcal{U}\times \mathcal{V} \to \mathbb{R}$, assume that $h(u,V)$ is $\gamma^2(u)$-sub-Gaussian under $P_{V}$ for all $u\in\mathcal{U}$ and $h(U,V)$ is $\frac{\sigma^2}{4}$-sub-Gaussian under both $P_{U}P_{V}$ and $P_{U,V}$. Assume also that $\log\mathbb{E}_{P_{U^*}}[e^{\gamma^2(U^*)}]<\infty$. Then for $\alpha\in(0,1)$,
    \begin{align*}
    \left|\mathbb{E}_{P_{U,V}}[h(U,V)] - \mathbb{E}_{P_U P_V}[h(U,V)]\right|\leq \sqrt{2((1-\alpha)\sigma^2+\alpha\log\mathbb{E}_{P_{U^*}}[e^{\gamma^2(U^*)}]) \, \frac{I_{\alpha}^{S}(P_{U,V},P_{U^*})}{\alpha}}.
    \end{align*}
\end{lemma}
\smallskip\noindent
{\bf Proof.}
    We first show that \( h(U,V) - \mathbb{E}_{P_V}[h(U,V)] \) is \(\sigma^2\)-sub-Gaussian under \(P_{U,V}\). By assumption, \(h(U,V)\) is \(\frac{\sigma^2}{4}\)-sub-Gaussian under \(P_{U,V}\). It remains to show that the term \(- \mathbb{E}_{P_V}[h(U,V)]\) is also \(\frac{\sigma^2}{4}\)-sub-Gaussian under \(P_{U,V}\).
Observe that
\begin{align}
\mathbb{E}_{P_{U,V}}\left[e^{\lambda \left(\mathbb{E}_{P_V}[h(U,V)] - \mathbb{E}_{P_{U,V}}\left[\mathbb{E}_{P_V}[h(U,V)]\right] \right)}\right] 
&= \mathbb{E}_{P_U}\left[e^{\lambda \left(\mathbb{E}_{P_V}[h(U,V)] - \mathbb{E}_{P_U P_V}[h(U,V)] \right)}\right] \nonumber \\
&\leq \mathbb{E}_{P_U P_V}\left[e^{\lambda \left(h(U,V) - \mathbb{E}_{P_U P_V}[h(U,V)] \right)}\right] \label{ASse1} \\
&\leq \exp\left(\frac{\lambda^2 \sigma^2}{8}\right), \label{ASse2}
\end{align}
where~\eqref{ASse1} follows from Jensen's inequality, and~\eqref{ASse2} follows from the assumption that \(h(U,V)\) is \(\frac{\sigma^2}{4}\)-sub-Gaussian under \(P_U P_V\). Thus, \(\mathbb{E}_{P_V}[h(U,V)]\) is \(\frac{\sigma^2}{4}\)-sub-Gaussian under \(P_{U,V}\).

Therefore, by Lemma~\ref{SubAdd}, it follows that \(-\mathbb{E}_{P_V}[h(U,V)]\) is \(\frac{\sigma^2}{4}\)-sub-Gaussian under \(P_{U,V}\), and hence the difference \(h(U,V) - \mathbb{E}_{P_V}[h(U,V)]\) is \(\sigma^2\)-sub-Gaussian under \(P_{U,V}\), as claimed.

    \smallskip
    Let $g:\mathcal{U}\times \mathcal{V} \to \mathbb{R}$ be defined by $g(u,v)= h(u,v) - \mathbb{E}_{P_V}[h(u,V)]$. Since, $h(U,V) - \mathbb{E}_{P_V}[h(U,V)]$ is $\sigma^2$-sub-Gaussian under $P_{U,V}$ and by definition of $g$, we obtain that for all $\lambda\in\mathbb{R}$:
    \begin{align*}        \log\mathbb{E}_{P_{U,V}}\left[e^{\left({\lambda\left( g(U,V) - \mathbb{E}_{P_{U,V}}[g(U,V)]\right)}\right)}\right]\leq \frac{\sigma^2\lambda^2}{2}.
    \end{align*}
    This can be re-written as
    \begin{align}\label{PUV}
    \log\mathbb{E}_{P_{U,V}}\left[e^{\left({\lambda\left( g(U,V)\right)}\right)}\right]\leq \frac{\sigma^2\lambda^2}{2}  + \mathbb{E}_{P_{U,V}}[\lambda g(U,V)].
    \end{align}
    Moreover, since, $h(u,V)$ is $\gamma(u)^2$-sub-Gaussian under $P_{V}$ for all $u\in\mathcal{U}$, it follows that for all $\lambda'\in\mathbb{R}$, 
    \begin{align}\label{Sub-gaus1}    \log\mathbb{E}_{P_{V}}\left[e^{\left({\lambda'\left( h(u,V) - \mathbb{E}_{P_{V}}[h(u,V)]\right)}\right)}\right]\leq \frac{\gamma^2(u)\lambda'^2}{2}.
    \end{align}
    Using the definition of $g$ and applying the exponential to both sides, we obtain: 
    \begin{align*}    \mathbb{E}_{P_{V}}\left[e^{\left({\lambda'\left( g(u,V) \right)}\right)}\right]\leq e^{\frac{\gamma^2(u)\lambda'^2}{2}}.
    \end{align*}
    Taking the expectation with respect to $P_{U^*}$ yields
    \begin{align*}    \mathbb{E}_{P_{U^*}P_{V}}\left[e^{\left({\lambda'\left( g(U^*,V) \right)}\right)}\right]\leq \mathbb{E}_{P_{U^*}}\bigg[e^{\frac{\gamma^2(U^*)\lambda'^2}{2}}\bigg].
    \end{align*}
    Finally, taking the logarithm and noting that $$\mathbb{E}_{P_{U^*} P_V }[g(U^*,V)] = \mathbb{E}_{P_{U^*} P_V }[h(U^*,V)] - \mathbb{E}_{P_{U^*} P_V }[h(U^*,V)] = 0,$$ we conclude that 
    \begin{align}\label{PU^*PV}        \log\mathbb{E}_{P_{U^*}P_{V}}\left[\exp\left({\lambda'\left( g(U^*,V)\right)}\right)\right]\leq \frac{\lambda'^2}{2}\log\mathbb{E}_{P_{U^*}}[e^{\gamma^2(U^*)}]  + \mathbb{E}_{P_{U^*}P_{V}}[\lambda'g(U^*,V)].
    \end{align}
    Using the Donsker-Varadhan representation for $D_{\text{KL}}(P_{\hat{U},\hat{V}}\|P_{U,V})$~\cite{DV} and inequality \eqref{PUV}, we have for all $\lambda\in\mathbb{R}$ that 
    \begin{align*}
       D_{\text{KL}}(P_{\hat{U},\hat{V}}\|P_{U,V}) &\geq \mathbb{E}_{P_{\hat{U},\hat{V}}}[\lambda g(\hat{U},\hat{V})] - \log\mathbb{E}_{P_{U,V}}[e^{\lambda g(U,V)}]  \\
       &\geq \mathbb{E}_{P_{\hat{U},\hat{V}}}[\lambda g(\hat{U},\hat{V})] - \mathbb{E}_{P_{U,V}}[\lambda g(U,V)] \mbox{} - \frac{\sigma^2\lambda^2}{2}.
    \end{align*}
    Rearranging terms yields 
    \begin{align}\label{PEq1}
        \lambda\left(\mathbb{E}_{P_{\hat{U},\hat{V}}}[g(\hat{U},\hat{V})] - \mathbb{E}_{P_{U,V}}[g(U,V)]\right) \leq D_{\text{KL}}(P_{\hat{U},\hat{V}}\|P_{U,V})+\frac{\sigma^2\lambda^2}{2}.
    \end{align}
    
\smallskip

    \noindent Note that  $$\mathbb{E}_{P_{U,V}}[g(U,V)] = \mathbb{E}_{P_{U,V}}[\left(h(U,V)-\mathbb{E}_{P_V}[h(U,V)]\right)] = \mathbb{E}_{P_{U,V}}[h(U,V)] - \mathbb{E}_{P_U P_V}[h(U,V)]. $$

\smallskip

\noindent Similarly, using the Donsker-Varadhan representation for $D_{\text{KL}}(P_{\hat{U},\hat{V}}\|P_{U^*}P_{V})$~\cite{DV} and inequality \eqref{PU^*PV}, we have for all $\lambda'\in\mathbb{R}$ :
    \begin{align}\label{PEq2}
        \lambda'\left(\mathbb{E}_{P_{\hat{U},\hat{V}}}[g(\hat{U},\hat{V})] - \mathbb{E}_{P_{U^*}P_{V}}[g(U^*,V)]\right) \leq D_{\text{KL}}(P_{\hat{U},\hat{V}}\|P_{U^*}P_{V})+\frac{\lambda'^2}{2}\log\mathbb{E}_{P_{U^*}}[e^{\gamma^2(U^*)}].
    \end{align}

\smallskip

\noindent If $\lambda>0$, then choosing $\lambda'=\frac{\alpha}{\alpha-1}\lambda<0$, we have from \eqref{PEq1} that
 \begin{align}\label{Eq1}
        \mathbb{E}_{P_{\hat{U},\hat{V}}}[g(\hat{U},\hat{V})] - \mathbb{E}_{P_{U,V}}[g(U,V)]\leq \frac{D_{\text{KL}}(P_{\hat{U},\hat{V}}\|P_{U,V})}{\lambda}+\frac{\sigma^2\lambda}{2}.
 \end{align}
 On the other hand,  since $\mathbb{E}_{P_{U^*} P_V }[g(U^*,V)] = 0, $ \eqref{PEq2} yields with $\lambda' = \frac{\alpha}{\alpha-1}\lambda<0$:
\begin{align}\label{Eq2}
        -\mathbb{E}_{P_{\hat{U},\hat{V}}}[g(\hat{U},\hat{V})] \leq \frac{D_{\text{KL}}(P_{\hat{U},\hat{V}}\|P_{U^*}P_{V})}{|\lambda'|}+\frac{|\lambda'|}{2}\log\mathbb{E}_{P_{U^*}}[e^{\gamma^2(U^*)}].
    \end{align}

\noindent Adding \eqref{Eq1} and \eqref{Eq2} yields that for all $\lambda>0$:
\begin{align}\label{BEq1}
    -\mathbb{E}_{P_{U,V}}[g(U,V)]  &\leq \frac{D_{\text{KL}}(P_{\hat{U},\hat{V}}\|P_{U,V})}{\lambda}+\frac{\sigma^2\lambda}{2} + \frac{D_{\text{KL}}(P_{\hat{U},\hat{V}}\|P_{U^*}P_{V})}{|\lambda'|} +\frac{|\lambda'|}{2}\log\mathbb{E}_{P_{U^*}}[e^{\gamma^2(U^*)}] \nonumber \\
    &= \frac{D_{\text{KL}}(P_{\hat{U},\hat{V}}\|P_{U,V})}{\lambda}+\frac{\sigma^2\lambda}{2} + \frac{D_{\text{KL}}(P_{\hat{U},\hat{V}}\|P_{U^*}P_{V})}{\frac{\alpha}{1-\alpha}\lambda} \nonumber \\
    &\hspace{0.5cm}+\frac{\frac{\alpha}{1-\alpha}\lambda}{2}\log\mathbb{E}_{P_{U^*}}[e^{\gamma^2(U^*)}] \nonumber \\
    &= \frac{\alpha D_{\text{KL}}(P_{\hat{U},\hat{V}}\|P_{U,V})+(1-\alpha)D_{\text{KL}}(P_{\hat{U},\hat{V}}\|P_{U^*}P_{V})}{\alpha\lambda} \nonumber \\ &\hspace{0.5cm} +   \lambda\left(\frac{(1-\alpha)\sigma^2+\alpha\log\mathbb{E}_{P_{U^*}}[e^{\gamma^2(U^*)}]}{2(1-\alpha)}\right).
\end{align}    

\noindent Similarly for $\lambda<0$, choosing $\lambda'=\frac{\alpha}{\alpha-1}\lambda>0$, we have from ~\eqref{PEq1} and ~\eqref{PEq2} that
\begin{align}\label{Eq3}
        -\left(\mathbb{E}_{P_{\hat{U},\hat{V}}}[g(\hat{U},\hat{V})] - \mathbb{E}_{P_{U,V}}[g(U,V)]\right)\leq \frac{D_{\text{KL}}(P_{\hat{U},\hat{V}}\|P_{U,V})}{|\lambda|}+\frac{\sigma^2|\lambda|}{2}.
    \end{align}
and 
\begin{align}\label{Eq4}
        \mathbb{E}_{P_{\hat{U},\hat{V}}}[g(U,V)] \leq \frac{D_{\text{KL}}(P_{\hat{U},\hat{V}}\|P_{U^*}P_{V})}{\lambda'}+\frac{\lambda'}{2}\log\mathbb{E}_{P_{U^*}}[e^{\gamma^2(U^*)}].
    \end{align}
Adding \eqref{Eq3} and \eqref{Eq4} yields for all $\lambda<0$ and $\lambda' = \frac{\alpha}{\alpha-1}\lambda>0$ that
\begin{align}\label{BEq2}
     \mathbb{E}_{P_{U,V}}[g(U,V)]  &\leq \frac{D_{\text{KL}}(P_{\hat{U},\hat{V}}\|P_{U,V})}{|\lambda|}+\frac{\sigma^2|\lambda|}{2} + \frac{D_{\text{KL}}(P_{\hat{U},\hat{V}}\|P_{U^*}P_{V})}{\lambda'}+\frac{\lambda'}{2}\log\mathbb{E}_{P_{U^*}}[e^{\gamma^2(U^*)}] \nonumber \\
    &= \frac{D_{\text{KL}}(P_{\hat{U},\hat{V}}\|P_{V,U})}{|\lambda|}+\frac{\sigma^2|\lambda|}{2} + \frac{D_{\text{KL}}(P_{\hat{U},\hat{V}}\|P_{U^*}P_{V})}{\frac{\alpha}{1-\alpha}|\lambda|}\nonumber \\*
    &\hspace{0.5cm}+\frac{\frac{\alpha}{1-\alpha}|\lambda|}{2}\log\mathbb{E}_{P_{U^*}}[e^{\gamma^2(U^*)}] \nonumber \\
    &= \frac{\alpha D_{\text{KL}}(P_{\hat{U},\hat{V}}\|P_{V,U})+(1-\alpha)D_{\text{KL}}(P_{\hat{U},\hat{V}}\|P_{U^*}P_{V})}{\alpha|\lambda|} \nonumber \\&\hspace{0.5cm} + |\lambda|\left(\frac{(1-\alpha)\sigma^2+\alpha\log\mathbb{E}_{P_{U^*}}[e^{\gamma^2(U^*)}]}{2(1-\alpha)}\right). 
\end{align}
\noindent Considering \eqref{BEq1} and \eqref{BEq2}, we have a non-negative parabola in $\lambda$ given by
\begin{align*}
    \lambda^2\left(\frac{(1-\alpha)\sigma^2+\alpha\log\mathbb{E}_{P_{U^*}}[e^{\gamma^2(U^*)}]}{2(1-\alpha)}\right) - \lambda\left(\mathbb{E}_{P_{U,V}}[h(U,V)] - \mathbb{E}_{P_U P_V}[h(U,V)]\right) \\ +\frac{\alpha D_{\text{KL}}(P_{\hat{U},\hat{V}}\|P_{U,V})+(1-\alpha)D_{\text{KL}}(P_{\hat{U},\hat{V}}\|P_{U^*}P_{V})}{\alpha} \geq 0,
\end{align*}
whose discriminant must be non-positive (for $\lambda=0$, the above inequality holds trivially). Thus for all $\alpha\in(0,1)$,
\begin{align}
    \MoveEqLeft\left|\mathbb{E}_{P_{U,V}}[h(U,V)] - \mathbb{E}_{P_U P_V}[h(U,V)]\right| &\nonumber \\ & \leq \sqrt{2\frac{((1-\alpha)\sigma^2+\alpha\log\mathbb{E}_{P_{U^*}}[e^{\gamma^2(U^*)}])}{(1-\alpha)}}\nonumber \\
    &\hspace{1cm}\times\sqrt{\left(\frac{\alpha D_{\text{KL}}(P_{\hat{U},\hat{V}}\|P_{U,V})+(1-\alpha)D_{\text{KL}}(P_{\hat{U},\hat{V}}\|P_{U^*}P_{V})}{\alpha}\right)}.
\end{align}
Finally, invoking Lemma~\ref{Asiblem} we obtain 
\begin{align}
    \left|\mathbb{E}_{P_{U,V}}[h(U,V)] - \mathbb{E}_{P_U P_V}[h(U,V)]\right| &\leq \sqrt{2((1-\alpha)\sigma^2+\alpha\log\mathbb{E}_{P_{U^*}}[e^{\gamma^2(U^*)}]) \, \frac{I_{\alpha}^{S}(P_{U,V},P_{U^*})}{\alpha}}.
\end{align}
\qed

\medskip
\noindent We next use Lemma 3.2 to derive our upper bound on the excess minimum risk in terms of the Sibson mutual information.

\begin{theorem}\label{Mtheorem3}
Let $X$, $Y$ and $Z$ be random vectors such that $Y\rightarrow X \rightarrow Z$ form a Markov chain as described in Section~\ref{subsec:problem_setup}. Assume that there exists an optimal estimator $f$ of Y from X such that $l(y,f(X))$ is conditionally $\gamma^2(y)$-sub-Gaussian under $P_{X|Z}$ for all $y\in\mathbb{R}^p$, where $\log\mathbb{E}_{P_ZP_{Y^*|Z}}[e^{\gamma^2(Y^*)}]<\infty$, and $l(Y,f(X))$ is conditionally $\frac{\sigma^2}{4}$-sub-Gaussian under both $P_{Y|Z}P_{X|Z}$ and $P_{Y,X|Z}$ , i.e., for all $\lambda\in\mathbb{R}$
\begin{align*}
    \log\mathbb{E}_{P_{X|Z}}\left[e^{(\lambda(l(y,f(X)))-\mathbb{E}_{P_{X|Z}}[l(y,f(X))])}\right]\leq \frac{\gamma^2(y)\lambda^2}{2} 
\end{align*}
    for all $y\in\mathbb{R}^p$,    $$\log\mathbb{E}_{P_{Y|Z}P_{X|Z}}\left[e^{(\lambda(l(Y,f(X)))-\mathbb{E}_{P_{Y|Z}P_{X|Z}}[l(Y,f(X))])}\right]\leq \frac{\sigma^2\lambda^2}{8}$$ and 
    $$\log\mathbb{E}_{P_{Y,X|Z}}\left[e^{(\lambda(l(Y,f(X)))-\mathbb{E}_{P_{Y,X|Z}}[l(Y,f(X))])}\right]\leq \frac{\sigma^2\lambda^2}{8}.$$ Then for $\alpha\in(0,1)$, the excess minimum risk satisfies
    \begin{align}\label{LT1AS}
        L_{l}^{*}(Y|Z)-L_{l}^{*}(Y|X) \leq\sqrt{\frac{2((1-\alpha)\sigma^2+\alpha\mathbb{E}_{P_Z}[\Phi_{Y^*|Z}(\gamma^2(Y^*))])}{\alpha} \, \mathbb{E}_{P_Z}\left[I_{\alpha}^{S}(P_{Y,X|Z},P_{Y^*|Z})\right]}, 
    \end{align}
 where $\Phi_{P_U}(V) = \log\mathbb{E}_{P_U}[e^{V}]$  and the distribution  $P_{Y^*|Z}$ has density 
\begin{align}\label{LT1 : subeq}
     p_{Y^*|Z}(y|z) = \frac{\left(\displaystyle\int \left(\frac{p_{Y,X|Z}(y,x|z)}{p_{Y|Z}(y|z)p_{X|Z}(x|z)}\right)^\alpha p_{X|Z}(x|z)dx\right)^{\frac{1}{\alpha}}}{\displaystyle\int\left(\int\left(\frac{p_{Y,X|Z}(y',x'|z)}{p_{Y|Z}(y'|z)p_{X|Z}(x'|z)}\right)^\alpha p_{X|Z}(x'|z) dx'\right)^{\frac{1}{\alpha}}p_{Y|Z}(y'|z)dy'}p_{Y|Z}(y|z).
 \end{align}  
\end{theorem}
\smallskip\noindent
{\bf Proof.}
    Let $\bar{X}$, $\bar{Y}$ and $\bar{Z}$ be the same random variables as defined in Theorem~\ref{Mtheorem}. We also consider the distribution $P_{Y^*|Z}$ given by $p_{Y^*|Z}(y|z)$ in \eqref{LT1 : subeq} obtained from  Definition~\ref{ASibM2} by considering the distributions $P_{Y,X|Z}$ and $P_{Y|Z}P_{X|Z}$.

    We apply Lemma~\ref{DecoupAS} by setting $U=Y$, $V=X$ and $h(u,v)=l(y,f(x))$ and taking regular expectations with respect to $P_{Y,X|Z=z}$ and $P_{\bar{Y},\bar{X}|Z=z}$. Since, $\bar{Y}$ and $\bar{X}$ are conditionally independent given $\bar{Z}=z$ such that $P_{\bar{Y},\bar{X}|Z=z} = P_{Y|Z}P_{X|Z}$, and $P_{\bar{Z}} = P_{Z}$, we have that
    \begin{align}\label{lemmaAS}
        \MoveEqLeft[6]|\mathbb{E}_{P_{Y,X|Z=z}}[l(Y,f(X))]-\mathbb{E}_{P_{\bar{Y},\bar{X}|Z=z}}[l(\bar{Y},f(\bar{X}))]| \nonumber \\ &= |\mathbb{E}_{P_{Y,X|Z=z}}[l(Y,f(X))]-\mathbb{E}_{P_{Y|Z=z}P_{X|Z=z}}[l(Y,f(X))]| \nonumber\\
        &\leq\sqrt{\frac{2((1-\alpha)\sigma^2+\alpha\Phi_{Y^*|Z=z}(\gamma^2(Y^*))) \,}{\alpha}I_{\alpha}^{S}(P_{Y,X|Z=z},P_{Y^*|Z=z})}.
    \end{align}
Now,
    \begin{align}
        &\left|\mathbb{E}_{P_{Y,X}}[l(Y,f(X))]- \mathbb{E}_{P_{\bar{Y},\bar{X}}}[l(\bar{Y},f(\bar{X}))]\right| \nonumber \\ &  = \left|\mathbb{E}_{P_Z}\left[\mathbb{E}_{P_{Y,X|Z=z}}[l(Y,f(X))]-\mathbb{E}_{P_{\bar{Y},\bar{X}|Z=z}}[l(\bar{Y},f(\bar{X}))]\right]\right|\nonumber \\
        & \leq \mathbb{E}_{P_Z}\left[\left|\mathbb{E}_{P_{Y,X|Z=z}}[l(Y,f(X))]-\mathbb{E}_{P_{\bar{Y},\bar{X}|Z=z}}[l(\bar{Y},f(\bar{X}))])\right|\right] \nonumber \\
        &\leq \mathbb{E}_{P_Z}\Bigg[\sqrt{2((1-\alpha)\sigma^2+\alpha\Phi_{Y^*|Z=z}(\gamma^2(Y^*))) \,\frac{I_{\alpha}^{S}(P_{Y,X|Z=z},P_{Y^*|Z=z})}{\alpha}}\, \Bigg] \nonumber \\
        & \leq\sqrt{2\mathbb{E}_{P_Z}\left[((1-\alpha)\sigma^2+\alpha\Phi_{Y^*|Z=z}(\gamma^2(Y^*)))\right]\mathbb{E}_{P_Z}\left[\frac{I_{\alpha}^{S}(P_{Y,X|Z=z},P_{Y^*|Z=z})}{\alpha}\right]} \nonumber 
        \\
        & = \sqrt{\frac{2((1-\alpha)\sigma^2+\alpha\mathbb{E}_{P_Z}[\Phi_{Y^*|Z}(\gamma^2(Y^*))])}{\alpha} \, \mathbb{E}_{P_Z}\left[I_{\alpha}^{S}(P_{Y,X|Z},P_{Y^*|Z})\right]},\label{TheoremASM}
    \end{align}
    where the first equality holds since $P_{\bar{Z}} = P_{Z}$, the first inequality  follows from Jensen's inequality, the second inequality follows from~\eqref{lemmaAS}, and the third inequality follows from the Cauchy-Schwarz inequality. From the proof of Theorem~\ref{Mtheorem}, we know that $\bar{Y}\rightarrow \bar{Z} \rightarrow \bar{X}$ forms a Markov chain. Hence,
    \begin{align}
        \mathbb{E}_{P_{\bar{Y},\bar{X}}}[l(\bar{Y},f(\bar{X}))])\geq L_{l}^{*}(Y|Z). \label{TI1AS}
    \end{align}
    Since $f$ is an optimal estimator of $Y$ from $X$, we also have
    \begin{equation}\label{TI2AS}
        \mathbb{E}_{P_{Y,X}}[l(Y,f(X))])= L_{l}^{*}(Y|X).
    \end{equation}
    Therefore using~\eqref{TI1AS} and~\eqref{TI2AS} in~\eqref{TheoremASM} along with the fact that $L_{l}^{*}(Y|Z)\geq L_{l}^{*}(Y|X)$, we arrive at the desired inequality:
    \begin{align*}
        \MoveEqLeft[2]L_{l}^{*}(Y|Z)-L_{l}^{*}(Y|X) & \nonumber \\*&\leq\sqrt{\frac{2((1-\alpha)\sigma^2+\alpha\mathbb{E}_{P_Z}[\Phi_{Y^*|Z}(\gamma^2(Y^*))])}{\alpha} \, \mathbb{E}_{P_Z}\left[I_{\alpha}^{S}(P_{Y,X|Z},P_{Y^*|Z})\right]}.
    \end{align*}
\qed

\medskip

Setting $\gamma^2(Y^*)=\sigma^2$ and taking the limit as $\alpha \to 1$ on the right-hand side of~\eqref{LT1AS} recovers the mutual information based bound~\eqref{MIB} of~\cite[Theorem~3]{Gyorfi} in the case of a constant sub-Gaussian parameter. 
We conclude this section by presenting a specialization of Theorem~\ref{Mtheorem3} to the case of bounded loss functions.

\smallskip

\begin{corollary}\label{CorASM}
    Suppose the loss function $l$ is bounded. Then for random vectors $X$, $Y$ and $Z$ such that $Y\rightarrow X \rightarrow Z$ as described in Section~\ref{subsec:problem_setup}, we have the following inequality for $\alpha\in(0,1)$ on the excess minimum risk:
    \begin{align}\label{CR1ASM}
        L_{l}^{*}(Y|Z)-L_{l}^{*}(Y|X)\leq\frac{\|l\|_{\infty}}{\sqrt{2}}\sqrt{\frac{(4-3\alpha)}{\alpha} \, \mathbb{E}_{P_Z}\left[I_{\alpha}^{S}(P_{Y,X|Z},P_{Y^*|Z})\right]}.
    \end{align}
\end{corollary}

\smallskip

\section{Numerical Results}
\smallskip

In this section, we present three examples where some of the proposed information divergence based bounds outperform the mutual information based bound. The first example considers a concatenated $q$-ary symmetric channel with a bounded loss function. The remaining two examples involve Gaussian additive noise channels and loss functions with non-constant sub-Gaussian parameters.

\begin{figure*}[htb]
    \centering
    \begin{subfigure}[t]{0.495\textwidth}
        \centering
        \includegraphics[width=\textwidth]{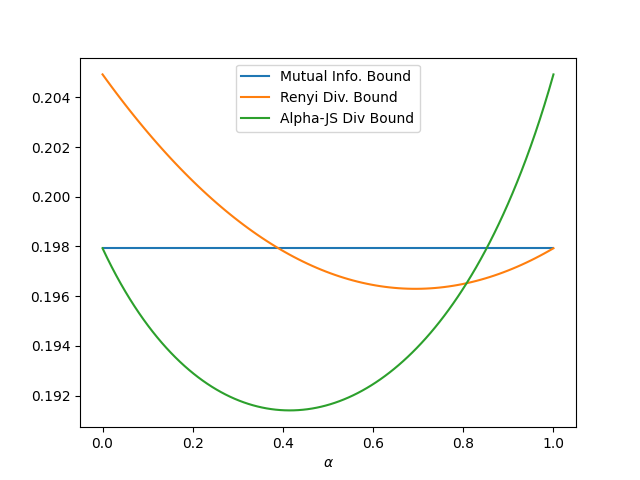}
        \caption{$q=2$ with $p = [0.3,0.7]$.}
        \label{fig:bsc-ex}
    \end{subfigure}
    \hfill
    \begin{subfigure}[t]{0.495\textwidth}
        \centering
        \includegraphics[width=\textwidth]{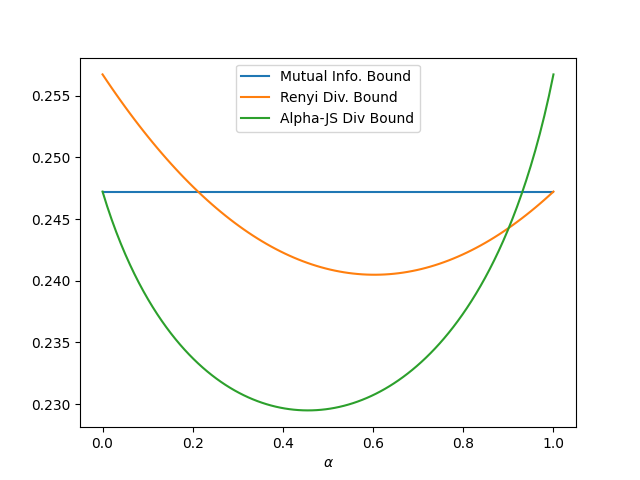}
        \caption{$q=3$ with $p = [0.4,0.2,0.4]$.}
        \label{fig:3sc-ex}
    \end{subfigure}
    
    \vspace{0.5cm}
    
    \begin{subfigure}[t]{0.495\textwidth}
        \centering
        \includegraphics[width=\textwidth]{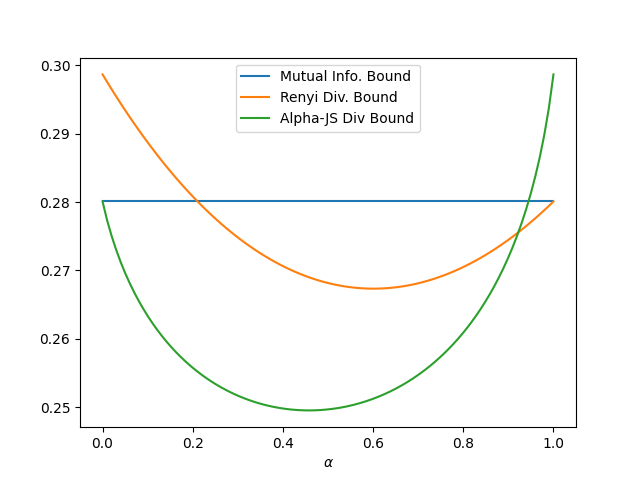}
        \caption{$q=5$ with $p = [0.25,0.1,0.4,0.15,0.1]$.}
        \label{fig:5sc-ex}
    \end{subfigure}
    \hfill
    \begin{subfigure}[t]{0.495\textwidth}
        \centering
        \includegraphics[width=\textwidth]{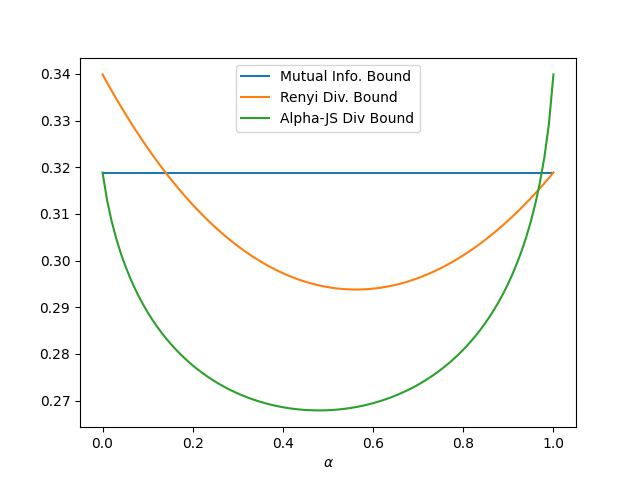}
        \caption{$q=10$, with $p$ drawn randomly.}
        \label{fig:10sc-ex}
    \end{subfigure}
    
    \vspace{0.5cm}
    
    \begin{subfigure}[t]{0.495\textwidth}
        \centering
        \includegraphics[width=\textwidth]{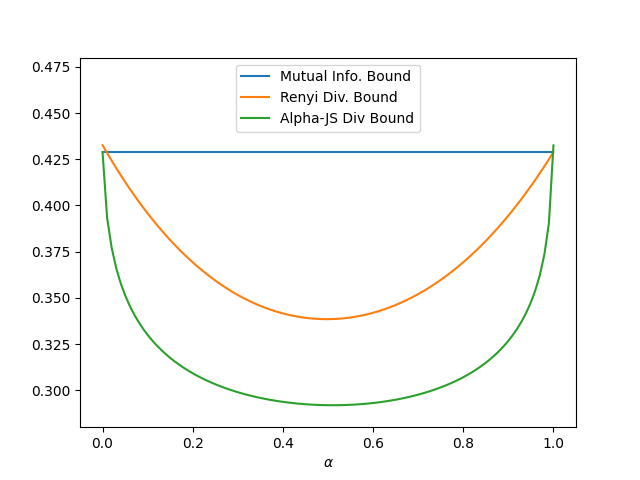}
        \caption{$q=100$, with $p$ drawn randomly.}
        \label{fig:100sc-ex}
    \end{subfigure}
    \hfill
    \begin{subfigure}[t]{0.495\textwidth}
        \centering
        \includegraphics[width=\textwidth]{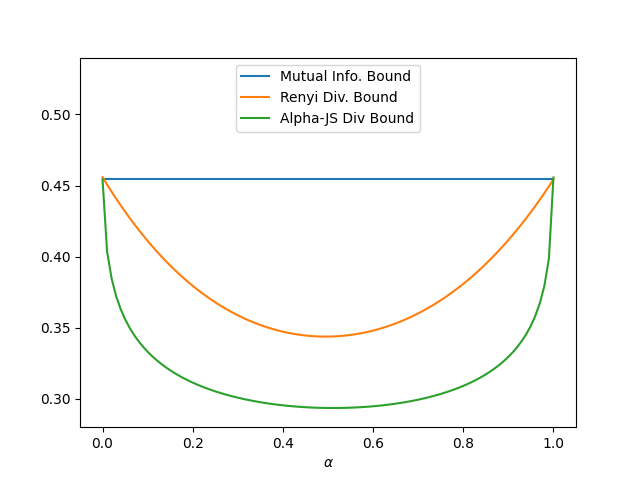}
        \caption{$q=200$, with $p$ drawn randomly.}
        \label{fig:200sc-ex}
    \end{subfigure}
    
    \caption{Comparison of bounds versus $\alpha$ on minimum excess risk for two concatenated $q$-ary symmetric channels, where $\epsilon_1=0.15$ and $\epsilon_2=0.05$.}
    \label{fig:qsc-ex}
\end{figure*}

\begin{example}
We consider a concatenation of two $q$-ary symmetric channels, with input $Y$ and noise variables $U_1$ and $U_2$, all taking values in $\{0, 1, \dots, q-1\}$. We assume that $Y,U_1$ and $U_2$  are independent. The input $Y$ has distribution $p = [p_0, p_1, \dots, p_{q-1}]$, while the noise variables $U_1$ and $U_2$ are governed by
$\mathbb{P}(U_i = 0) = 1 - \epsilon_i$ and $\mathbb{P}(U_i = a) = \epsilon_i / (q - 1)$
for all $a \in \{1, \dots, q - 1\}$ and $i=1,2$, where $\epsilon_1, \epsilon_2 \in (0,1)$ are crossover probabilities. The output $X$ of the first channel is given by 
$X = (Y + U_1) \bmod q$
and serves as the input to the second channel. The final output $Z$ is then given by $Z = (X + U_2) \bmod q,$ 
which can also be written as $Z = (Y + U_1 + U_2) \bmod q$. This construction naturally induces the Markov chain $Y \rightarrow X \rightarrow Z$.

\smallskip

Using a $0$--$1$ loss function (defined as $l(y, y') = 1(y \ne y')$, where $1(\cdot)$ denotes the indicator function), we compute the bounds in Corollary~\ref{CorR} and~\ref{CorJS}, corresponding to equations~\eqref{CR1} and~\eqref{CR1JS}, respectively, as functions of $\alpha \in (0,1)$. Figure~\ref{fig:qsc-ex} compares the R\'enyi-based bound~\eqref{CR1} and the $\alpha$-Jensen-Shannon based bound~\eqref{CR1JS} with the mutual information based bound~\eqref{MIB}. Among the two, the $\alpha$-Jensen-Shannon based bound consistently performs the best over a wide range of $\alpha$ values. Moreover, as $q$ increases in the $q$-ary symmetric channel, both the interval of $\alpha$ for which the proposed bounds outperform the mutual information based bound and the magnitude of improvement become more pronounced. 
For this example, we set $\epsilon_1 = 0.15$ and $\epsilon_2 = 0.05$. For $q = 10, 100, 200$, we generate input distributions by sampling from a symmetric (i.e., with identical parameters) Dirichlet distribution on $\mathbb{R}^q$. Using a Dirichlet parameter greater than one gives balanced distributions that avoid placing too much weight on any single symbol. For $q = 2, 3, 5$, the input distributions are explicitly specified in the figure captions.

Finally, we note that in this example, the specialized bound for bounded loss functions derived from the Sibson mutual information in Corollary~\ref{CorASM} does not offer any improvement over the standard mutual information based bound~\eqref{MIB} and is therefore not presented. In the next two examples, we compare the $\alpha$-Jensen-Shannon based bound of Theorem~\ref{MtheoremJS} with the mutual information based bound for loss functions with non-constant sub-Gaussian parameters.

\end{example}

\begin{figure}[!ht]
\centering
\includegraphics[width=4in]{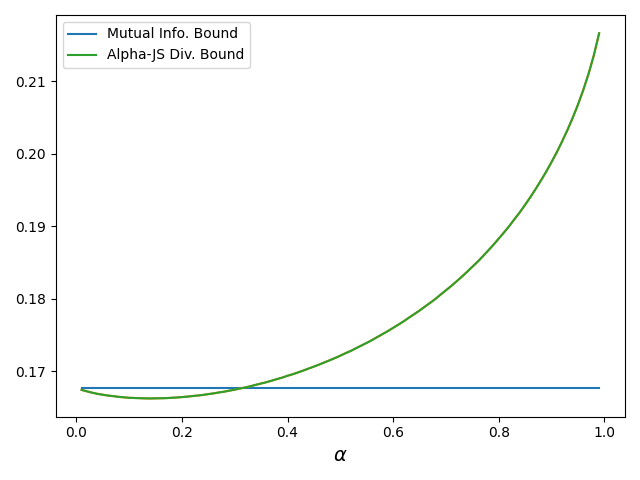}
\caption{Comparison of bounds vs $\alpha$ on minimum excess risk for a Gaussian additive noise channel with $c=1$, $\hat{\sigma}^2 = 1$  and  $\sigma^2_{i} = 1$ for all $i=1,2$.}\label{fig-AGN-ex}
\end{figure}

\begin{example}
    Consider a Gaussian additive noise channel with input $Y$ and noise random variables $W_{1}$ and $W_{2}$, where $Y \sim \mathcal{N}(0, \hat{\sigma}^2)$, $W_1 \sim \mathcal{N}(0, \sigma_1^2)$, and $W_2 \sim \mathcal{N}(0, \sigma_2^2)$. Assume that $Y$ is independent of $(W_1, W_2)$ and $W_1$ is independent of  $W_2$. Define $$X = Y + W_1$$ and $$Z = X + W_2 = Y + W_1 + W_2,$$ inducing the Markov chain $Y \rightarrow X \rightarrow Z$.

\smallskip

We consider the loss function $l(y, y') = \min\{|y - y'|, |y - c|\}$ for some $c > 0$. For this model, we observe that
\begin{align*}
l(y, f^*(X)) &= \min\{|y - f^*(X)|, |y - c|\} 
\leq |y - c| 
 \leq |y| + |c| 
= |y| + c,
\end{align*}
where $f^*$ denotes the optimal estimator of $Y$ from $X$. Thus, $l(y, f^*(X))$ is a non-negative random variable that is almost surely bounded by $|y| + c$. By Hoeffding’s lemma, it follows that this loss is conditionally $\sigma^2(y)$-sub-Gaussian under $P_{X|Z}$, $P_{X|Z,Y=y}$ and $P^{(\alpha)}_{X|Z,Y=y}$ for all $y \in \mathbb{R}^p$, with
$$
\sigma^2(y) = \frac{(|y| + c)^2}{4}.
$$

\smallskip

\noindent Furthermore, for $\hat{\sigma}^2=1$ and $\sigma^2_{i} = 1$, $i=1,2$, we have that
\begin{align*}
    \mathbb{E}[\sigma^2(Y)] &= \mathbb{E}\left[\frac{(|Y| + c)^2}{4}\right] 
    = \frac{1}{4} \mathbb{E}\left[|Y|^2 + c^2 + 2|Y|c\right] 
    = \frac{1 + c^2 + 2c \sqrt{2/\pi}}{4}.
\end{align*}

\smallskip

Hence, the conditions of Theorem~\ref{MtheoremJS} are satisfied. Figure~\ref{fig-AGN-ex} compares the $\alpha$-Jensen-Shannon based bound in~\eqref{LT1JS} with the mutual information based bound in~\eqref{MIB} for $c=1$. We observe that the $\alpha$-Jensen-Shannon based bound is tighter for values of $\alpha$ approximately in the range $(0, 0.3)$.
\end{example}

\begin{figure}[!ht]
\centering
\includegraphics[width=4in]{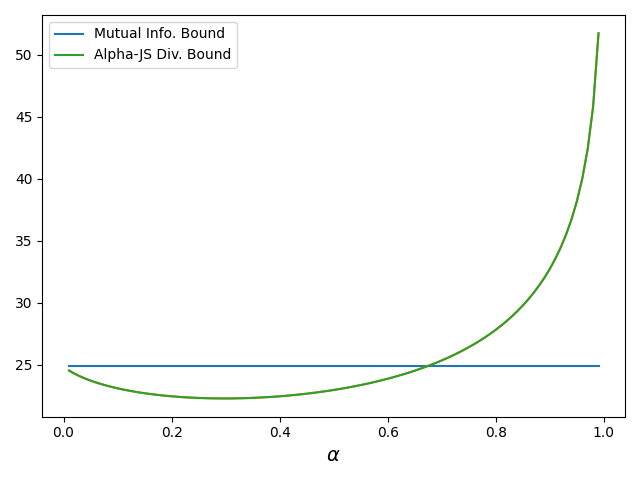}
\caption{Comparison of bounds vs $\alpha$ on minimum excess risk for a reverse Gaussian additive noise channel with $c=1$, $\hat{\sigma}^2 = 2$,  $\sigma_1^2 = 39$ and $\sigma_2^2 = 1$.}\label{fig-RAGN-ex}
\end{figure}

\begin{example}
Consider a Gaussian additive noise model with input $Z \sim \mathcal{N}(0, \hat{\sigma}^2)$ and two noise variables $W_1 \sim \mathcal{N}(0, \sigma_1^2)$ and $W_2 \sim \mathcal{N}(0, \sigma_2^2)$, all mutually independent. Let 
$$X = Z + W_1$$ 
and 
$$Y = X + W_2 = Z + W_1 + W_2,$$ 
inducing the Markov chain $Z \rightarrow X \rightarrow Y$, which is equivalent to the Markov chain $Y \rightarrow X \rightarrow Z$.

\smallskip

We again consider the loss function $l(y, y') = \min\{|y - y'|, |y - c|\}$ for some $c > 0$, and observe that $l(y, f^*(X)) \leq |y| + c$, where $f^*$ is the optimal estimator of $Y$ from $X$. Hence, the loss is (conditionally) $\sigma^2(y)$-sub-Gaussian as in the previous example with $\sigma^2(y) = \frac{(|y| + c)^2}{4}$. For $\hat{\sigma}^2 = 2$, $  \sigma_1^2 = 39$ and $\sigma_2^2 = 1$, the expected sub-Gaussian parameter is
\begin{equation*}
\mathbb{E}[\sigma^2(Y)] = \frac{42 + c^2 + 2c\sqrt{84/\pi}}{4}.
\end{equation*}
Therefore, the conditions of Theorem~\ref{MtheoremJS} continue to hold.

\smallskip

In contrast to the previous example, where $Y$ was the input and $Z$ the degraded observation, this example reverses that direction. Figure~\ref{fig-RAGN-ex} compares the $\alpha$-Jensen-Shannon based bound in~\eqref{LT1JS} with the mutual information based bound in~\eqref{MIB} for $c=1$. We observe that the $\alpha$-Jensen-Shannon based bound is tighter for values of $\alpha$ approximately in the range $(0, 0.7)$.
\end{example}

\clearpage



\section{Conclusion}\label{sec:Conclusion}
\smallskip

In this paper, we studied the problem of bounding the excess minimum risk in statistical inference using generalized information divergence measures. Our results extend the mutual information based bound in~\cite{Gyorfi} by developing a family of bounds parameterized by the order $\alpha \in (0,1)$, involving the R\'enyi divergence, the $\alpha$-Jensen-Shannon divergence, and Sibson’s mutual information. For the R\'enyi divergence based bounds, we employed the variational representation of the divergence, following the approach in~\cite{Modak}, and for the $\alpha$-Jensen-Shannon and Sibson-based bounds, we adopted the auxiliary distribution method introduced in~\cite{Aminian}.

\smallskip

Unlike the bounds in~\cite{Modak,Aminian}, which assume the sub-Gaussian parameter to be constant, our framework allows this parameter to depend on the (target) random vector, thereby making the bounds applicable to a broader class of joint distributions. We demonstrated the effectiveness of our approach through three numerical examples: one involving concatenated discrete $q$-ary symmetric channels, and two based on additive Gaussian noise channels. In all cases, we observed that at least one of our $\alpha$-parametric bounds is tighter than the mutual information based bound over certain ranges of $\alpha$, with the improvements becoming more pronounced in the discrete example as the channel alphabet size $q$ increased.

\smallskip

Future directions include exploring bounds under alternative $f$-divergence measures, developing tighter bounds for high-dimensional settings, and determining divergence rates in infinite-dimensional cases.

\bibliographystyle{IEEEtran}
\bibliography{citations}

\end{document}